\documentclass{emulateapj}
\slugcomment{{\sc Accepted to ApJ:} August 8, 2011}

\renewcommand{\deg}{\mbox{$^{\circ}$}}
\newcommand{\kms}{\mbox{km s$^{-1}$}}

\newcommand{\uJy}{\mbox{$\mu$Jy}}

\newcommand{\x}{\mbox{$\times$}}

\newcommand{\E}[1]{\hbox{$10^{ #1 }$}}

\newcommand{\about}{\mbox{$\sim$}}

\begin{document}


\title{Resolution of the compact radio continuum sources in A\lowercase{rp}220}

\author{Fabien Batejat, John E. Conway and  Rossa Hurley} \affil{Onsala Space Observatory, SE-439 92 Onsala, Sweden}
\email{\tt fabien.batejat@chalmers.se}

\author{Rodrigo Parra} \affil{European Southern Observatory, Alonso de Cordova 3107, Casilla 19001, Santiago 19, Chile}

\author{Philip J. Diamond} \affil{CSIRO Astronomy and Space Science, P.O. Box 76, Epping, NSW 1710, Australia}

\author{Colin J. Lonsdale} \affil{MIT Haystack Observatory, Westford MA, USA}

\and\author{Carol J. Lonsdale} \affil{North American ALMA Science Center, NRAO, Charlottesville, VA, USA}


\begin{abstract}

We present 2~cm and 3.6~cm wavelength very long baseline interferometry images of the  compact radio continuum 
sources in the nearby ultra-luminous infrared galaxy Arp220. Based on their radio spectra and  variability properties, we confirm these sources 
to be  a mixture of supernovae (SNe) and supernova remnants (SNRs).  Of the 
17 detected sources we resolve 7 at both wavelengths. The 
SNe generally only have upper size limits. In contrast all the SNRs are 
resolved with diameters $\geq0.27~\mathrm{pc}$. This size limit is consistent with them having just entered their 
Sedov phase while embedded in an interstellar medium (ISM) of density $10^{4}~\mathrm{cm^{-3}}$.  These objects lie 
on the diameter--luminosity correlation for SNRs  (and so also on the diameter--surface brightness 
relation) and extend these correlations to very small sources.  The data are consistent  with the relation  
$L~\propto D^{-9/4}$.
Revised equipartition arguments adjusted to a magnetic field to relativistic particle energy density ratio of 1\% combined with a reasonable synchrotron-emitting 
volume filling factor of 10\% give estimated magnetic field strengths in the SNR shells of $\sim15$--$50~\mathrm{mG}$.
The SNR shell magnetic fields are unlikely to come from compression of ambient ISM fields and must instead be internally generated.
We set an upper limit of 7~mG for the ISM magnetic field.
The estimated energy in relativistic particles, 2\%--20\% of the explosion kinetic energy,
is consistent with estimates from models that fit the IR--radio correlation in compact starburst galaxies.

\end{abstract}



\section{Introduction}\label{se:introduction}

Observations of radio supernovae (SNe) and supernova remnants (SNRs)  provide an important means to study astrophysical processes 
occurring in dense nuclear starbursts. Radio observations of these objects are
free from the dust obscuration that hamper observations at shorter wavelengths while the high angular resolution afforded by very long baseline interferometry (VLBI) observations cannot be matched by any other technique. 
Monitoring the rate of appearance of radio SNe can potentially constrain the stellar initial mass function (IMF) and check whether it is modified in extremely dense environments such as found in Arp220 \citep{PARRA07}.
The SNRs are the acceleration sites of the relativistic particles that give rise to radio emission in star-forming regions and hence their properties are central to understanding the FIR--radio correlation \citep{LACKI10}.
Finally, radio SNe and SNRs can be used as in situ probes to constrain the interstellar medium (ISM) properties such as density, pressure, and magnetic field strength. Recent papers dealing with VLBI observations of SNe/SNRs in starburst galaxies include \citet{FENECH10}, \citet{LENC09}, \citet{ULVESTAD09}, and \citet{PEREZ09}.

 \begin{deluxetable*}{cccccc}
\centering
\tablecolumns{6}
\tabletypesize{\scriptsize}
\tablewidth{0pt}

\tablecaption{High Frequency VLBI Observations of Arp220}

\tablehead{
\colhead{Epoch}&
\colhead{Code}&
\colhead{$\lambda$}&
\colhead{Array}&
\colhead{$\sigma_\mathrm{rms}$}&
\colhead{Beam Size}
\\
\colhead{}&
\colhead{}&
\colhead{(cm)}&
\colhead{}&
\colhead{(\uJy~$\mathrm{beam}^{-1}$)}&
\colhead{(mas$^{2}$)}}
\startdata
2006.02&BP129  &13.26  & VLBA     &129.54 & 3.6\x 6.6\\
           &       & 6.02$^\mathrm{a}$  &          &41.43 & 1.8\x 3.5\\
           &       & 3.56  &          &86.73 & 1.7\x 3.1\\
2006.91 & GC028A & 3.56 & Global VLBI & 34.7& 0.56\x1.34\\
2006.99	& GC028B & 2  & HSA  & 28.12 & 0.43\x1.26\\
2008.44 & GC031A & 6.02 & Global VLBI & 12.58 & 0.72\x2.1 \\
	   
\enddata
\tablecomments{$^\mathrm{a}$ Re-reduced 6~cm observations from experiment BP129.}
\label{ta:observations}
\end{deluxetable*}

Arp220 is the nearest and best studied ultra-luminous infrared galaxy (ULIRG). For over a decade it has been the subject of a global VLBI
campaign at cm wavelengths. \citet{SMITH98} made the first detection of a number of compact sub-parsec sized sources at 
18~cm. These sources have been  monitored and new objects discovered in  successive 18~cm  epochs \citep{ROVILOS05,LONSDALE06}.
\citet{PARRA07} reported the first detections of these objects at the shorter wavelengths of  13~cm, 6~cm, and 3.6~cm. 
Based on  the source radio  light-curves and spectra, \citet{PARRA07} argued that the compact radio sources in Arp220 comprise a mixed
population of radio SNe and SNRs. 

This paper presents  the results and analysis of  new high sensitivity and resolution VLBI observations at 3.6~cm and 2~cm wavelengths. 
The main  objectives of these new observations were to extend source spectra to higher frequency,
look for high-frequency variability, and most importantly attempt to spatially resolve sources or set limits on source sizes. 
 In Section~\ref{se:reduction} we present details of the high-frequency observations and their reduction;  we  also describe other 
ancillary VLBI data used in our analysis.  In Section~\ref{se:results} we describe our results including the source detection and resolution criteria, the estimation of 
source sizes and of source spectra. In Section~\ref{se:discussion} we discuss our results and their implications for  the physics of the sources and the ambient 
ISM. Finally in Section~\ref{se:conclusions} we give  our conclusions. In this paper we assume a distance of 77~Mpc to Arp220 at which an angular size of 1~mas corresponds to 0.37~pc.


\section{Observations and data reduction}\label{se:reduction}

\subsection{3.6~cm Global VLBI }\label{se:reduction_x}

Arp220 was observed at 3.6~cm (8.4~GHz) on 2006 November 28 as part of European VLBI Network (EVN) + Very Long Baseline Array (VLBA) global VLBI experiment GC028A.
The experiment used the 10 stations of the VLBA, the Green Bank Telescope (Gb),  the phased Very Large Array (VLA; Y27), Arecibo (Ar), and
five EVN stations (Ef, On, Mc, Nt, Wb). The observations were performed using a dual polarization $256~\mathrm{Mbit~s^{-1}}$ recording mode. 
The data were correlated at JIVE in the Netherlands in a single pass using a phase center between the two nuclei and high spectral and time 
resolution to allow wide-field imaging. Unfortunately, no fringes were detected at the Green Bank Telescope which significantly reduced
the sensitivity of the observations. Also due to a scheduling error at the station Ar was only available for a very small fraction of the 
time.  

The data were reduced using AIPS in a standard manner. Initial amplitude calibration was carried out using monitored system temperatures and station gains. We estimate that absolute amplitude calibration should be accurate to \about5\%.
The main phase calibrator was the nearby ($\mathrm{0\rlap.^{\circ}5}$ separation) source 
 J1532+2344 which was observed in a rapid switching cycle with Arp220. 
Simultaneous CLEAN images of size 4096~pixels $\times$ 2048~pixels (pixel spacing 0.1~mas)  were then made of the eastern and western nuclei using 
robust data weighting to give a good compromise between sensitivity and resolution. 
Using the initial phase-referencing phase calibration it was found  impossible to make thermal noise limited images. To improve 
 the dynamic range we performed phase self-calibration on the Arp220 data itself using long solution times 
 ($\approx$ 30 minutes) to obtain sufficient signal to noise. The resulting slowly varying phase solutions were consistent 
 with  unmodeled atmosphere delays differentially affecting target and calibrator. Finally CLEAN was 
 applied again and noise limited images produced (see  Table~\ref{ta:observations}). The final images are shown in the central panels of Figures~\ref{fig:west_map} and \ref{fig:east_map}, 
with inset panels showing detailed images of individual sources. In the rest of this paper we refer to experiment GC028A 
when mentioning the 3.6~cm data or image if not otherwise specified.

\subsection{2~cm HSA}\label{se:reduction_u}

\begin{figure*}[htp]
  \centering
   \includegraphics[width=0.85\hsize]{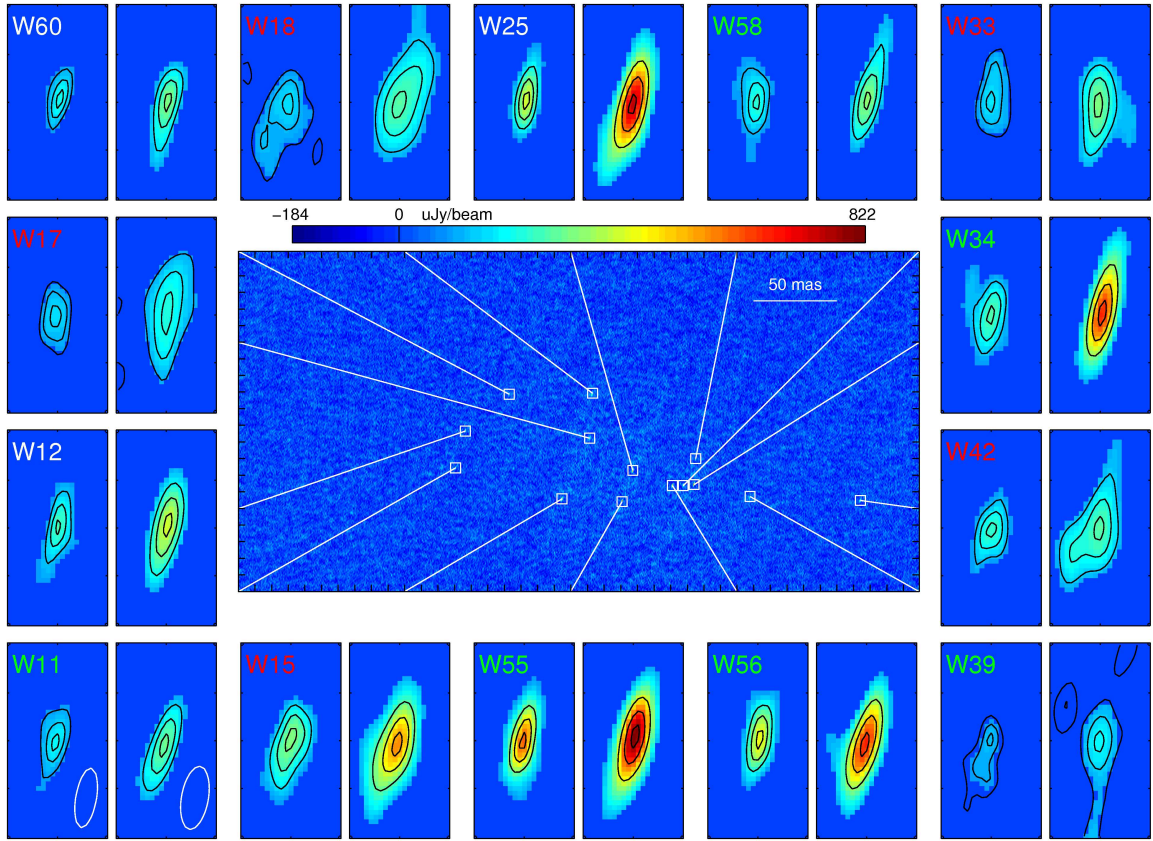}
     \caption{Central panel shows the 3.6~cm image of the western nucleus of Arp220 spanning a region 410~mas by 205~mas. Fourteen sources were detected above a
     5.6$\sigma$ detection threshold (see Section~\ref{se:detection}) at both 2~cm and 3.6~cm.
     The surrounding panels show zoomed images of these sources at 2~cm (left panels) and at 3.6~cm (right panels). 
     The same color bar is used for the central panel and all zoomed images at both 2~cm and 3.6~cm.
      Each zoomed image  covers a region of 2.1~mas by 4.1~mas. The source number is displayed in the top left corner of the 2~cm
     zoomed image. Where the source number is displayed in red this means that the source is resolved at both
     2~cm and 3.6~cm. A green number means resolved at 2~cm only, a cyan number means resolved at 3.6~cm only while a white number means unresolved at both 2~cm and 3.6~cm. For more information about how resolution was determined see Section~\ref{se:resolution}. The zoomed images are blanked at the 3$\sigma$ level and contour levels
     are shown at 50\%, 75\%, and 95\% of the peak intensity of each source (given in Table~\ref{ta:source_fluxes}). 
     The 50\% contour of the CLEAN restoring beam (see Table~\ref{ta:observations}) at each wavelength is shown in the bottom left 
      zoomed image.}
  \label{fig:west_map}
  \end{figure*}

\begin{figure*}[htp]
  \centering
   \includegraphics[width=0.85\hsize]{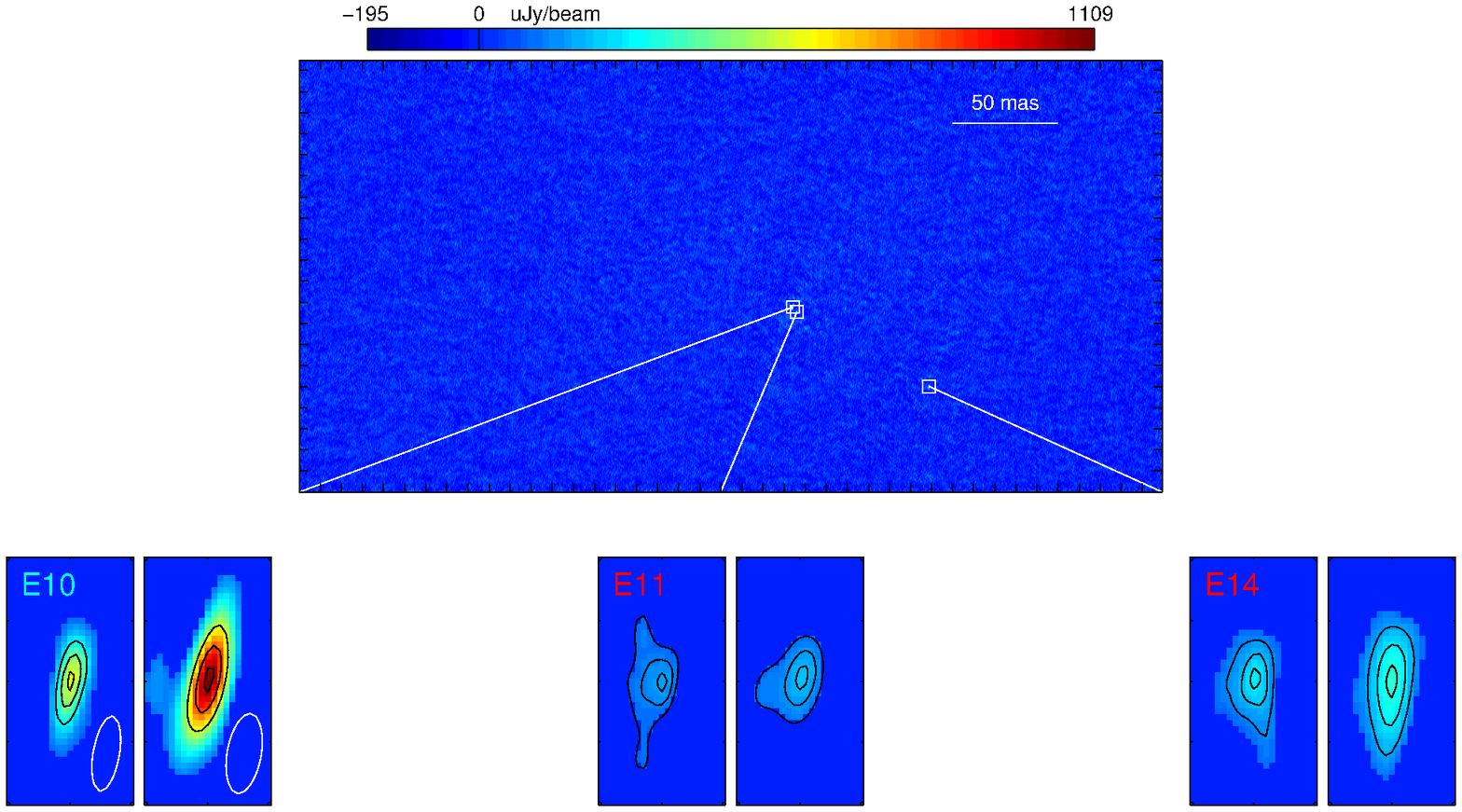}
     \caption{Central panel shows the 3.6~cm image of the eastern nucleus of Arp220 spanning a region 410~mas by 205~mas. Three sources were detected above a
     5.6$\sigma$ detection threshold (see Section~\ref{se:detection}) at both 2~cm and 3.6~cm. For information on labeling, contours, and other image properties see Figure~\ref{fig:west_map}.
     }
  \label{fig:east_map}
  \end{figure*}

Arp220 was observed at  2~cm (15~GHz) in a full track observation on 2006 December 28 using the 
High Sensitivity Array (HSA; expt GC028B).
The experiment used the 10 stations of the VLBA, the Green Bank Telescope (Gb), the phased VLA (Y27), and
 the Effelsberg (Eb) antenna.  The observations were performed in a  $256~\mathrm{Mbit~s^{-1}}$ dual polarization mode. 
The data were correlated in Socorro, New Mexico in one pass with a single phase center located midway 
between the two nuclei and high spectral and time resolution to allow wide-field imaging.
The data were reduced using AIPS  in a standard way  similar to that described in
 Section~\ref{se:reduction_x}. After experimentation with many uv data weighting schemes pure natural weighting of the
 visibility  data was chosen. This choice gave the highest possible sensitivity at the expense of resolution; however  for other choices most sources
 were not detectable.   As for our 3.6~cm wavelength data, phase self-calibration of the Arp220  data was required to achieve a  thermal noise
limited image  (see Table~\ref{ta:observations}).  Our resulting 2~cm wavelength images for individual sources in each nucleus are shown in insets in Figures~\ref{fig:west_map} and \ref{fig:east_map}, respectively.

\subsection{Other VLBI Data}\label{se:reduction_other}

A number of ancillary VLBI data sets, providing  
multifrequency monitoring information, were also used in the analysis presented in this
paper. These observations were used 
to primarily classify sources as SNe or SNRs (see Section~\ref{se:source_id}).  A full presentation of 
these monitoring data, including data on many sources only detected at lower frequency, 
will be given in a future paper (Batejat et al. 2012, in preparation).

 The most extensive monitoring data exist at 18~cm wavelength where a total of nine epochs spanning from 1994.87 to 2006.43 
 have been observed and reduced. These data provide information on source long wavelength variability on decade timescales.
We have reanalyzed the complete 18~cm data set in a consistent fashion carefully 
correcting for epoch dependent variations in  flux scale.  Additionally to this, shorter wavelength variability can be  searched for
 by comparing the results  at  6~cm  from epoch 2006.02  \citep[experiment BP129, see][]{PARRA07}
re-reduced by us in order to achieve a lower $\sigma_{\mathrm{rms}}$ (see Table~\ref{ta:source_fluxes} for the new flux densities) with data  2.42~years later at epoch 2008.44 from a recently reduced observation
 (experiment GC031A, see Table~\ref{ta:observations}). This latter epoch is one of a series of  6~cm 
  monitoring experiments  which we are presently  reducing. Finally we compare our new GC028A 
 3.6~cm  data with data from \citet{PARRA07} taken 0.9 years earlier. Because of the low
 SNR of this old 3.6~cm wavelength data it has not been possible to re-reduce it using self-calibration techniques (unlike 6~cm data from the same experiment).

\section{Results and analysis}\label{se:results}

\subsection{Source Detection}\label{se:detection}

Visual inspection of the final 2~cm and 3.6~cm images (Figures~\ref{fig:west_map} and \ref{fig:east_map}) shows many 
sources clearly visible at both wavelengths.  To more rigorously  define a list of detected 
sources we applied a two-pass  search method as used by \citet{PARRA07}.
First  the whole east and west nucleus images were searched for sources significantly
above the noise, using a detection limit which minimized the chance of false detections. Then in a second 
pass small regions were searched around the positions of known sources (previously detected at 18~cm, 6~cm and 3.6~cm since the first detection of compact sources by \citet{SMITH98}) using a lower detection threshold. 
For a detection limit  set to  $\eta\sigma$ where $\sigma$ is the noise
rms, the probability of one or more false detections in an image is  $F=1-P(I<\eta\sigma)^{N_\mathrm{s}}$, 
where $P(I<\eta\sigma)$ is the cumulative
probability of a noise spike at a given beam area being less than
$\eta\sigma$ and  $N_\mathrm{s}$ is the number of beam areas searched. Histograms of the pixel values
in source free areas at both wavelengths were close to Gaussian out to the most extreme 
pixel values, hence $P$ could be  calculated assuming Gaussian statistics.  In both passes
we  chose $\eta$ such that $F$ was $0.2\%$. Given that we have two passes and two wavelengths, this 
choice gave a final probability of $<1\%$ of getting one or more false detections.  

For the first pass, given the number of beam areas at 2~cm, the above chosen value of $F$ corresponded 
to $\eta = 5.6$. Although the number of 
searched beams was slightly less at 3.6~cm wavelength we conservatively used the same
detection criteria. This first pass applied separately to both wavelengths resulted in the detection of 
14 sources in the western nucleus (two of which, namely  W58 and W60, were new detections)  and 3 in the 
eastern nucleus. All of these sources were detected at both wavelengths. 
In the second pass the search regions were limited to boxes of $4~\mathrm{mas^2}$ centered on the positions 
 of known sources. Given the  smaller number of beam areas searched in this pass the critical 
detection  threshold  was smaller ($\eta = 3.8$). Despite this lower limit no additional detections were 
made.

Table~\ref{ta:source_fluxes} gives the absolute  positions of the 17 detected sources, estimated by fitting a Gaussian to
the 2~cm image of each source. Given the relatively close-by calibrator source used ($0\rlap.^{\circ}5$ distant)
the  absolute  astrometric accuracy is likely limited  \citep{PRADEL06} by the accuracy of the phase calibrator position
\citep[estimated for J1532+234 to be 0.5~mas;][]{PETROV05}.
In Table~\ref{ta:source_fluxes} we therefore give positions rounded to the nearest milliarcsecond in each coordinate. Columns 9--12 of Table~\ref{ta:source_fluxes} give source peak and total  flux densities at 3.6~cm and 2~cm from experiments GC028A and B. The total flux densities were measured within  a tight box surrounding each source. For other epochs and wavelengths (Columns 5--8) only peak flux densities are given; since however sources are unresolved in these data, peak brightness measurements should also be good estimates of source total flux densities. Note that the 6~cm flux density values given for BP129 in Table~\ref{ta:source_fluxes} are based on a re-reduction of the data using self-calibration while the values at other wavelengths are taken from \citet{PARRA07}. For the re-reduced 6~cm data source flux densities are on average a  factor of 1.36 larger than found by \citet{PARRA07}; this increase is likely due to an increase in phase coherence of that data. A possible slight overestimate in flux densities due to noise biasing of low SNR self-calibrated data potentially could also be present but this potential effect is hard to quantify without detailed simulations.

\begin{deluxetable*}{cccccccccccc}

\centering
\tablecolumns{12}
\tabletypesize{\scriptsize}
\tablewidth{0pt}

\tablecaption{Position of Detected Radio Sources and Flux Densities}

\tablehead{
\colhead{}&
\colhead{}&
\colhead{$\alpha_{2000}$}&
\colhead{$\delta_{2000}$}&
\colhead{BP129}&
\colhead{BP129}&
\colhead{GC031A}&
\colhead{BP129}&
\colhead{GC028A}&
\colhead{GC028A}&
\colhead{GC028B}&
\colhead{GC028B}
\\
\colhead{Name}&
\colhead{SN}&
\colhead{$15^\mathrm{h}34^\mathrm{m}$...}&
\colhead{23\deg30\arcmin...}&
\colhead{13 cm}&
\colhead{6 cm}&
\colhead{6 cm}&
\colhead{3.6 cm}&
\colhead{3.6 cm}&
\colhead{3.6 cm}&
\colhead{2 cm}&
\colhead{2 cm}
\\
\colhead{}&
\colhead{}&
\colhead{}&
\colhead{}&
\colhead{Peak}&
\colhead{Peak}&
\colhead{Peak}&
\colhead{Peak}&
\colhead{Peak}&
\colhead{Integrated}&
\colhead{Peak}&
\colhead{Integrated}
\\
\colhead{(1)}&
\colhead{(2)}&
\colhead{(3)}&
\colhead{(4)}&
\colhead{(5)}&
\colhead{(6)}&
\colhead{(7)}&
\colhead{(8)}&
\colhead{(9)}&
\colhead{(10)}&
\colhead{(11)}&
\colhead{(12)}
}

\startdata
Error ($\sigma$) & & & & $\pm130$ & $\pm41$ & $\pm13$ & $\pm87$ & $\pm35$ & $\pm35$ & $\pm28$ & $\pm28$ \\
\hline
W11 & \nodata & 57\rlap.$^\mathrm{s}$2299 & 11\rlap.\arcsec502 & 573 & 581 & 818 & 357 & 289 & 279 & 190 & 163 \\
W12 & \nodata & 57\rlap.$^\mathrm{s}$2295 & 11\rlap.\arcsec524 & 1011 & 711 & 953 & 341 & 389 & 416 & 258 & 253 \\
W15 & \nodata & 57\rlap.$^\mathrm{s}$2253 & 11\rlap.\arcsec483 & 579 & 959 & 1115 & 707 & 546 & 837 & 314 & 500 \\
W17 & 6 & 57\rlap.$^\mathrm{s}$2241 & 11\rlap.\arcsec520 & 477 & 524 & 454 & 484 & 218 & 385 & 150 & 134 \\
W18 & 7 & 57\rlap.$^\mathrm{s}$2240 & 11\rlap.\arcsec547 & 559 & 834 & 738 & 451 & 272 & 560 & 156 & 288  \\
W25 & \nodata & 57\rlap.$^\mathrm{s}$2222 & 11\rlap.\arcsec501 & 1069 & 1141 & 1479 & 648 & 752 & 965 & 390 & 324 \\
W33 & 11 & 57\rlap.$^\mathrm{s}$2200 & 11\rlap.\arcsec491 & 582 & 417 & 321 & 397 & 310 & 409 & 162 & 182 \\
W34 & \nodata & 57\rlap.$^\mathrm{s}$2195 & 11\rlap.\arcsec492 & 699 & 1066 & 1458 & 743 & 650 & 758 & 263 & 358 \\
W39 & 12 & 57\rlap.$^\mathrm{s}$2171 & 11\rlap.\arcsec485 & 749 & 428 & 466 & 120 & 190 & 247 & 129 & 91 \\
W42 & 13 & 57\rlap.$^\mathrm{s}$2122 & 11\rlap.\arcsec482 & 743 & 693 & 706 & 574$^*$ & 273 & 492 & 240 & 251 \\
W55 & \nodata & 57\rlap.$^\mathrm{s}$2227 & 11\rlap.\arcsec482 & 124 & 1000 & 569 & 1147 & 822 & 1001 & 595 & 693 \\
W56 & \nodata & 57\rlap.$^\mathrm{s}$2205 & 11\rlap.\arcsec491 & \nodata & 905 & 1434 & 750 & 657 & 815 & 426 & 446 \\
W58 & \nodata & 57\rlap.$^\mathrm{s}$2194 & 11\rlap.\arcsec508 & \nodata & 408 & 723 & \nodata & 322 & 296 & 226 & 250 \\
W60 & \nodata & 57\rlap.$^\mathrm{s}$2276 & 11\rlap.\arcsec546 & \nodata & 285 & 1016 & \nodata & 310 & 268 & 200 & 128 \\
E10 & \nodata & 57\rlap.$^\mathrm{s}$2915 & 11\rlap.\arcsec335 & 227 & 1034 & 1631 & 988 & 1109 & 1394 & 577 & 596 \\
E11 & \nodata & 57\rlap.$^\mathrm{s}$2913 & 11\rlap.\arcsec333 & \nodata & 429 & 310 & \nodata & 201 & 171 & 165 & 195 \\
E14 & \nodata & 57\rlap.$^\mathrm{s}$2868 & 11\rlap.\arcsec298 & \nodata & 698 & 608 & 549 & 343 & 509 & 264 & 400 \\
\enddata
\tablecomments{Col.~(1):~source name from \citet{LONSDALE06} for all sources except
W55 and W56, which are from \citet{PARRA07}. Sources W58 and W60 are newly detected sources. W indicates the source is
located in the western nucleus while E stands for east.
Col.~(2):~Name used in \citet{SMITH98} and \citet{ROVILOS05}.
Cols.~(3) and (4):~J2000 Right Ascencion and Declination obtained by fitting a gaussian to the sources in
the highest frequency map (2~cm map).
Col.~(5):~13~cm peak flux density (in \uJy~$\mathrm{beam}^{-1}$) from the observations performed in experiment BP129 and
presented in \citet{PARRA07}. Col.~(6):~re-reduced 6~cm peak flux density (in \uJy~$\mathrm{beam}^{-1}$) from experiment BP129. Re-working of these data
allowed us to produce a deeper map of both nuclei with $\sigma_\mathrm{rms}=41.43~\uJy~\mathrm{beam}^{-1}$ and to detect two more sources in the western nucleus (W58 and W60) and one more source in the eastern nucleus (E11). Col.~(7):~Newly acquired (epoch 2008.44) 6~cm peak flux density (in \uJy~$\mathrm{beam}^{-1}$) which will be presented and
discussed in detail in Batejat et al. (2012) (in preparation). Col.~(8):~3.6~cm peak flux density (in \uJy~$\mathrm{beam}^{-1}$) from the observations performed in experiment BP129 and presented in \citet{PARRA07}. For source W42 (marked by an asterisk) we give the integrated flux density because this source appeared resolved. Cols.~(9) and (10):~3.6~cm peak and integrated
flux densities (respectively in \uJy~$\mathrm{beam}^{-1}$ and in \uJy) from experiment GC028A discussed in this paper. Cols.~(11) and (12):~2~cm peak and integrated
flux densities (respectively in \uJy~$\mathrm{beam}^{-1}$ and in \uJy) from experiment GC028B discussed in this paper.}
\label{ta:source_fluxes}
\end{deluxetable*}

\subsection{Size Estimation}\label{se:size_est}

A primary goal of our new high-frequency observations was to measure or set limits on the source sizes.
From  visual inspection of the inset images in  Figures~\ref{fig:west_map} 
and~\ref{fig:east_map}, there are  several candidate resolved sources (i.e., W18, W33, W42, and E14)  
where the 50\% of peak contour, at least at one wavelength, encloses a significantly  larger area  than the beam and so 
appear resolved. Quantitative  tests are however required to confirm  or reject these visual  impressions
and to  give size estimates with confidence limits.  Two methods were developed to do this. 
In the method described in  Section~\ref{se:resolution} we tested whether the null hypothesis that a source 
was  unresolved could be rejected while in Section~\ref{se:sizes} we  describe a procedure applied to give
best estimates (with error bars) of each source's  size. In Section~\ref{se:size_res} we summarize the results 
obtained after applying these methods to our data.

\subsubsection{Test of Resolution}\label{se:resolution}

We tested whether or not  each  source was consistent  with the null hypothesis of  a point source convolved with the 
CLEAN  Gaussian  restoring beam (having  minor and major axis dispersions of  $\sigma_{\mathrm{x}}$ and
$\sigma_{\mathrm{y}}$). 
For each source we calculated an observed  minor axis dispersion 
$ \sigma_{\mathrm{x_{obs}}}$  from
\begin{equation} \label{eq:dispersion}
\sigma_{\mathrm{x_{obs}}}^{2} = \frac{\int\int I(x,y) \Delta{x}^{2} dx dy}{\int\int  I(x,y) dx dy},
\end{equation}
\noindent where $\Delta x$ is measured along the minor axis through the source center. If, given 
the expected noise,  $\sigma_{\mathrm{x_{obs}}}$ was significantly larger than $\sigma_{\mathrm{x}}$ the point-source hypothesis
was rejected. Analytic calculation of how much the former quantity must exceed the latter to be confident 
of resolution is difficult, first because  of pixel-to-pixel correlations in noise,  and second because 
of blanking applied to the data.
Any practical method of estimating $\sigma_{\mathrm{x_{obs}}}$
must first blank the image $I(x,y)$ below say, 3$\sigma_\mathrm{rms}$ or else the statistic is
dominated by distant noise peaks. This blanking however makes the problem non-linear.

To circumvent the above  problems we applied a statistical bootstrapping technique. 
For each detected source a point of the same flux density convolved 
with the restoring beam was added at multiple (source-free) positions on the final image to test 
the effect of different realizations of the  noise.   Each realization was blanked below 3$\sigma_\mathrm{rms}$
and the  minor axis spatial dispersion  $\sigma_{\mathrm{x_{obs}}}$  measured. The resulting histogram of measured
dispersion values was then compared to that measured for
the source  using the same blanking criteria.
If this latter quantity was greater than 90\% of  the  histogram values the null hypothesis of a point source was 
provisionally rejected.

Table~\ref{ta:sim_results} lists for each source and wavelength the percentage probability of the point source  hypothesis
being rejected. According to our adopted criteria of 90\% confidence 13 sources are provisionally resolved at 
2~cm and  8 at 3.6~cm.  In total seven sources fulfill our  resolution criterion at both bands.  Since all  these  have $<1\%$ 
cumulative probability of  achieving their measured dispersion by chance if a point source we argue 
that for these sources the point source hypothesis can be rejected.

\subsubsection{Estimating  Source Sizes}\label{se:sizes}
 
Best estimates of source sizes with error bars were made using a variation on the method presented
in Section~\ref{se:resolution}. Again we added   a model source to the final image at multiple positions  
to take into account different realizations of the noise. In this case however the model added was 
 a spherical shell convolved with the CLEAN
 restoring beam,  whose  outer diameter we then adjusted at each position  till the measured  dispersion after blanking equaled the observed  source dispersion after blanking. At positions where even a source of zero size was unable to produce a minor axis dispersion small enough to equal that measured from the data we set the estimated source size to zero.
The SNR shell model used had a ratio of inner to outer diameter 0.8 similar to that measured for SN1993J \citep{MARTIVIDAL11} and SN2008iz \citep{BRUNTHALER10} and consistent with the observation of evolved SNRs in M82 \citep{KRON85,MUX94,BESWICK06,FENECH08,FENECH10}.
From the resulting histogram of  outer shell diameters  we then obtained  a best estimate of the source size and estimated error bars. 
 In some cases we obtained good fits for the outer diameter ($> 0$) in over 
 90\% of realizations. In these cases  we took the best estimate of source  size as the mean
  of the histogram over all realizations with failed solutions set to zero size and defined error bars from its  10th and 90th percentiles.
  In other cases with  $<90\%$ of realizations giving good solutions we still set a size upper limit at  the 90th percentile of fitted sizes but we set the lower limit to zero.

 \subsubsection{Summary of Results for Source Resolution/Size}\label{se:size_res}
 
\begin{figure*}[htp]
  \centering
   \includegraphics[width=0.9\hsize]{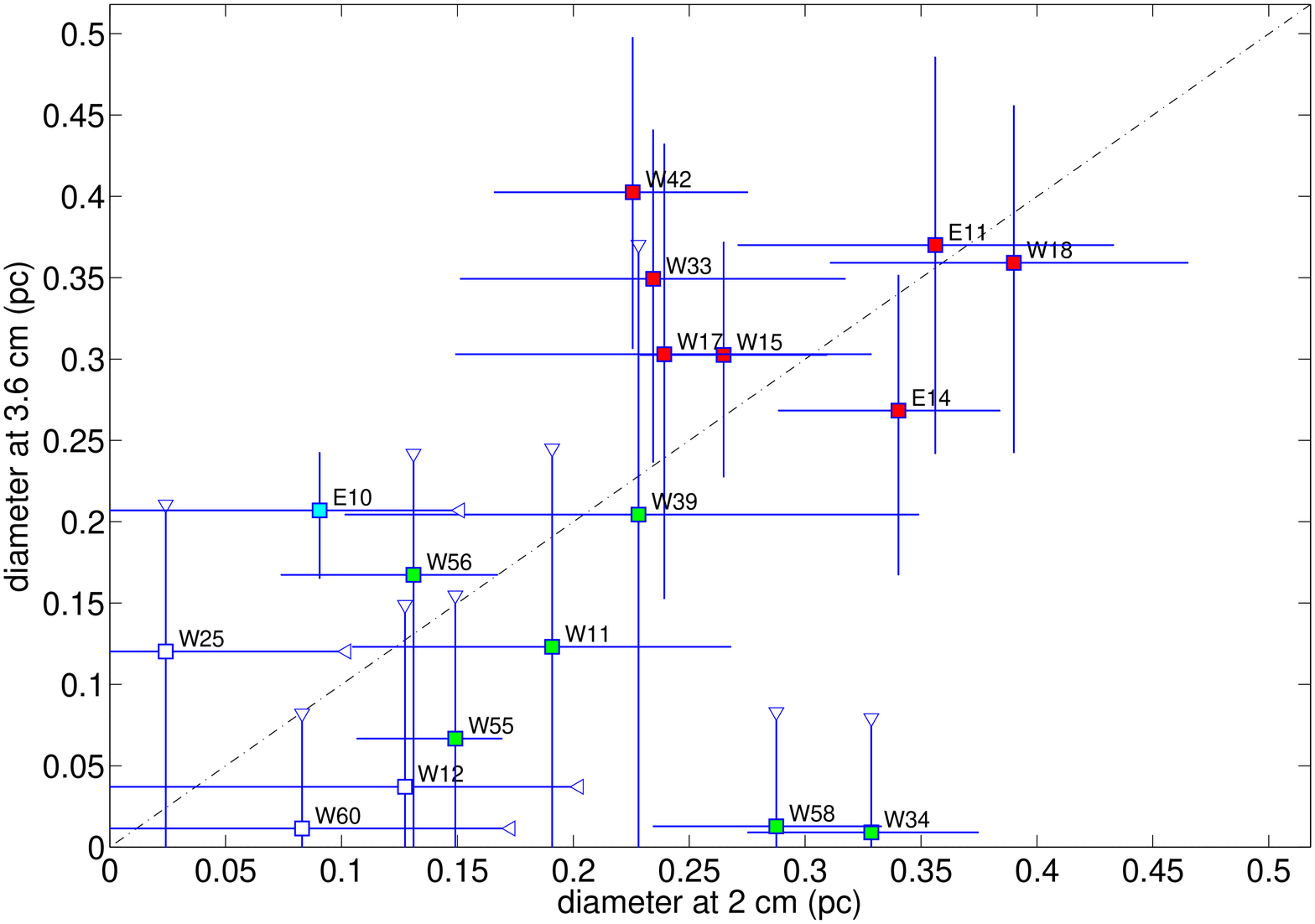}
  \caption{Plot of estimated source outer shell diameters  at 2~cm and 3.6~cm. The estimated source sizes and error
   bars at each wavelength are assigned as described in Section~\ref{se:sizes}. Symbol color indicates the 
  results of source resolution  testing described in Section~\ref{se:resolution}.  Red indicates that the
  source is resolved at both 2 and 3.6~cm defined as 99\% confidence rejection of the null
  hypothesis that the source is a point. Green means that the source is provisionally resolved at
  2~cm only and cyan that the source is provisionally resolved at 3.6~cm only. White represents
  sources that are unresolved at both bands.  
  }
  \label{fig:errorbar}
\end{figure*}

 The results of the statistical tests described in Sections~\ref{se:resolution} and \ref{se:sizes}
 are summarized in Figure~\ref{fig:errorbar}. 
 Three sources (W12, W25 and W60)  are  very compact and unresolved  at both wavelengths, using both resolution/size estimation methods.
 This demonstrates that residual atmospheric phase errors do not have any appreciable effect in broadening our sources, since such errors would affect  all sources.
 Seven  sources (E10, W11, W34, W39, W55, W56,  and W58)  are provisionally resolved at one wavelength 
 (usually 2~cm)  and a further seven sources (E11, E14, W15, W17, W18, W33, and W42) are resolved at both wavelengths. All of the latter group have 
 diameters greater than 0.7~mas.
 
 Figure~\ref{fig:errorbar} shows in general a strong correlation between 
 the sizes/limits  measured at the two bands. Two clear  exceptions however are W34 and W58, both of which are 
 well resolved at 2~cm but are unresolved at 3.6~cm.   Inspection of the detailed image for W34 at 2~cm shows 
 that an isolated probable noise feature to the north-west gets through the initial  $3\sigma$ blanking and contributes to a large minor axis
 spatial  dispersion.  The case of W58 is less clear since  there is no distinct peak off the main source just a broadened source with position 
 angle different to the CLEAN  beam. This could however be caused by a large noise peak that lies very close or on top of the 
 source. The presence of one or two  such anomalous sources  is not unexpected by chance given the blanking method used. There are 17 sources each at two 
frequencies with approximately 10 independent  beam areas per source box in Figures~\ref{fig:west_map} and \ref{fig:east_map}.  Given 
this total area there  is a probability of 37\% of detecting one or more such $>3\sigma$  noise peaks  and a 8\% chance of detecting 
two or more such peaks.

It should be noted that although well resolved at both bands both W42 and W33 seem to be  significantly larger at 3.6~cm  than 2~cm; possible explanations for this are discussed in Section~\ref{se:source_flat}. In addition the estimated size of W42 at 3.6~cm in GC028A (ring outer diameter 0.82~mas) is much less than  that claimed at the same frequency by  \citet{PARRA07} from earlier BP129 VLBA observations (equivalent to an outer shell  diameter of 3.6~mas after converting from the fitted Gaussian FWHM).  It should be noted that the beam area of these earlier VLBA observations was much 
larger, by a factor of six, compared to  our new global 3.6~cm observations. A consequence is that a shell of the size and brightness claimed from the BP129 observations would be below the thermal noise per beam of full resolution GC028A data  and so would be undetectable. To check this possibility lower resolution versions of the GC028A images and recent very sensitive global 6~cm observations (GC031A) were inspected but neither were consistent with the large shell structure 
claimed from the BP129 3.6~cm data.

The initial argument for the resolution of W42 at 3.6~cm in \citet{PARRA07} was based on having two large VLBA  beams  areas above 50\% of the peak source brightness. Significant residual atmospheric  phase errors in the data (expected to be larger than those known to be present at  6~cm) may have contributed to
an apparent source broadening,  unfortunately the SNR is too  low at 3.6~cm to allow self-calibration reprocessing of the BP129  to quantify atmospheric broadening at this wavelength. Residual phase errors at 3.6~cm  would have broadened all 
sources equally  but when combined with the low surface brightness of  W42 this might explain the BP129 result. From these data the 50\% of peak contour level is only three times the rms of the thermal  noise (see Figure 4 of \citet{PARRA07}). Given there were a total of 18 sources detected by  \citet{PARRA07} the probability of at least one of
them having a $3\sigma$ noise peak  adjacent to the source, so giving the impression of a highly resolved source,  is not negligible (2.5\%) and  furthermore this  probability of false resolution would increase rapidly in the presence of even a moderate amount of atmospheric source broadening.

 \subsection{Source Spectra}\label{se:spectra}

The short wavelength observations presented in this paper allow us to extend the source radio spectra first
studied by \citet{PARRA07} to higher frequency (15~GHz, 2~cm). In addition the 3.6~cm observations and new 6~cm observations (see Table~\ref{ta:observations})  can
be compared to  those in \citet{PARRA07} to look for high frequency variability. Both spectral  shapes and variability are useful
diagnostics when attempting to classify sources as SNRs or SNe (see the Appendix). Figure~\ref{fig:spectra} shows the source spectra with the points measured 
in the period 2006.02--2006.43 plotted in blue and those observed in the period 2006.91--2008.44 plotted in red.
The measured flux densities at each epoch are taken from Table~\ref{ta:source_fluxes}.


\section{Discussion} \label{se:discussion}

\begin{deluxetable*}{ccccccccccccc}
\centering
\tablecolumns{13}
\tabletypesize{\scriptsize}
\tablewidth{0pt}

\tablecaption{Source Resolution Probabilities and Size Estimates}

\tablehead{
\colhead{Source}&
\colhead{2 cm}&
\colhead{3.6 cm}&
\colhead{}&
\colhead{Radius 2 cm (mas)}&
\colhead{}&
\colhead{}&
\colhead{Radius 3.6 cm (mas)}&
\colhead{}&
\colhead{Radius}&
\colhead{Diameter}&
\colhead{Source ID}
\\
\cline{4-6}
\cline{7-9}
\colhead{}&
\colhead{(\%)}&
\colhead{(\%)}&
\colhead{Mean}&
\colhead{Min}&
\colhead{Max}&
\colhead{Mean}&
\colhead{Min}&
\colhead{Max}&
\colhead{(mas)}&
\colhead{(pc)}&
\colhead{}
\\
\colhead{(1)}&
\colhead{(2)}&
\colhead{(3)}&
\colhead{(4)}&
\colhead{(5)}&
\colhead{(6)}&
\colhead{(7)}&
\colhead{(8)}&
\colhead{(9)}&
\colhead{(10)}&
\colhead{(11)}&
\colhead{(12)}
}
\startdata
W11 & 95.2 & 68.8 & 0.258 & 0.142 & 0.362 & \nodata & \nodata & 0.331 & 0.26 &  0.19& SN\\
W12 & 88.4 & 24.8 & \nodata & \nodata & 0.274 & \nodata & \nodata & 0.201 & $<$0.28 &  $<$0.21 & T\\
W15 & 99.8 & 98.5 & 0.358 & 0.310 & 0.418 & 0.409 & 0.308 & 0.502 & \bf0.38 &  \bf0.28 & SNR\\
W17 & 96.3 & 90.5 & 0.324 & 0.202 & 0.444 & 0.410 & 0.207 & 0.584 & \bf0.37 &  \bf0.27& SNR\\
W18 & 99.9 & 98.1 & 0.527 & 0.421 & 0.629 & 0.486 & 0.328 & 0.616 & \bf0.51 &  \bf0.38& SNR\\
W25 & 33.0 & 72.4 & \nodata & \nodata & 0.138 & \nodata & \nodata & 0.285 & $<$0.29 &  $<$0.22 & T\\
W33 & 96.6 & 97.9 & 0.317 & 0.205 & 0.429 & 0.472 & 0.320 & 0.596 & \bf0.39 &  \bf0.29 & SNR\\
W34 & 99.9 & 4.8 & 0.444* & 0.372* & 0.506* & \nodata & \nodata & 0.100 & \nodata &  \nodata & SN\\
W39 & 92.7 & 69.3 & 0.309 & 0.138 & 0.472 & \nodata & \nodata & 0.501 & 0.31 &  0.23 & U\\
W42 & 98.2 & 99.1 & 0.305 & 0.225 & 0.372 & 0.544 & 0.415 & 0.672 & \bf0.42 &  \bf0.31 & SNR\\
W55 & 97.7 & 43.4 & 0.202 & 0.145 & 0.229 & \nodata & \nodata & 0.209 & 0.20 &  0.15 & SN\\
W56 & 92.5 & 87.0 & 0.177 & 0.100 & 0.226 & \nodata & \nodata & 0.327 & 0.18 &  0.13& SN\\
W58 & 99.8 & 10.9 & 0.389* & 0.317* & 0.450* & \nodata & \nodata & 0.101 & \nodata &  \nodata & SN\\
W60 & 64.3 & 9.8 & \nodata & \nodata & 0.234 & \nodata & \nodata & 0.100 & $<$0.24 & $<$0.18 & SN\\
E10 & 83.1 & 97.6 & \nodata & \nodata & 0.204 & 0.280 & 0.224 & 0.328 & 0.28 &  0.21& U\\
E11 & 99.5 & 97.4 & 0.482 & 0.367 & 0.585 & 0.500 & 0.327 & 0.656 & \bf0.49 &  \bf0.36 & SNR\\
E14 & 99.8 & 95.8 & 0.460 & 0.390 & 0.519 & 0.363 & 0.227 & 0.475 & \bf0.41 &  \bf0.30 & SN\\
\enddata
\tablecomments{Col. (1):~Source name from \citet{LONSDALE06} for all sources except
W55 and W56 which are from \citet{PARRA07}. Sources W58 and W60 are newly detected sources. Cols. (2) and (3):~probability ($P$)
of source being resolved at 2~cm and 3.6~cm, respectively, using the method described in Section~\ref{se:resolution}.
We classify a source as provisionally resolved at that frequency in cases $P\geq90\%$. Cols. (4)--(6):~the mean,
minimum, and maximum outer shell radii at 2~cm (experiment GC028B) as calculated by the algorithm described in
Section~\ref{se:sizes}. In cases where a source is unresolved only an upper limit is given, i.e., a max
value. Cols. (7)--(9):~the mean,
minimum, and maximum outer shell radii at 3.6~cm (experiment GC028A) as calculated by the algorithm described in
Section~\ref{se:sizes}. In cases where a source is unresolved only an upper limit is given, i.e., a max
value. Col. (10):~best estimate of outer shell radius. In cases where the source is resolved at both bands this
is calculated by averaging the values of Columns 4 + 7 in mas and the result is printed in bold font. Where a source is resolved only at one
band the radius is given as the mean value of the resolved band. Where a source is unresolved at
both bands, the radius is given as an upper limit taken as the bigger of the two max values. 
Col. (11):~outer shell diameter in parsecs. Col. (12):~source identification; SN = supernova, SNR = supernova remnant, T = transition
candidate, U = unclassified (see the Appendix). The * beside the 2~cm size estimates for W34 and W58 indicate that
we consider these unreliable (see Section~\ref{se:size_res}) and we therefore do not list a size for these sources in
Columns 10 and 11.}
\label{ta:sim_results}
\end{deluxetable*}

\subsection{Source Properties and Identification}\label{se:source_id}

\citet{PARRA07} argued that the compact radio sources in Arp220 comprise  a mixed population of SNe and 
SNRs; the former embedded in an ionized 
circumstellar bubble and  the latter strongly interacting with the surrounding ISM. 
 In addition to the above two classes   SN/SNR ``transition'' objects may also 
exist in which ISM interaction has begun  but the  swept up mass is less than the ejecta mass.  
Other compact sources might  be associated with active galactic nuclei (AGNs). Below we discuss the expected properties 
and likely members of each class. The Appendix gives a description of the variability 
and spectral properties for each source and its classification.

\vskip 0.2cm
\noindent{\it Supernovae.} In these objects the synchrotron-emitting blast wave transits the dense circumstellar medium 
of the progenitor star, originating from a pre-explosion stellar wind with a $r^{-2}$ density profile, which is ionized 
 by the SN explosion.  As the SN expands the competition between fading synchrotron
emission from the emitting shell and  decreasing free--free optical depth gives  a characteristic light-curve with  a  relatively fast 
rise followed by a  gradual decline, with the peak flux density occurring later  at longer wavelengths. Those of our 
sources which have had  rapid rises in flux density at 18~cm  and stable or decreasing flux densities at shorter wavelengths are almost 
certainly radio SNe. The expected radio spectra of such sources is a
power law with  a sharp cutoff toward long wavelengths. While 
foreground ionized  ISM  can also cause low-frequency turnovers in SNR sources these turnovers are expected only 
at frequencies $\leq 2~\mathrm{GHz}$ \citep{PARRA07}, hence sources with  turnovers at a higher frequency are most likely SNe.
Based on the above light-curve and spectrum criteria in the Appendix we provisionally  classify
W11, W34, W55, W56, W58, W60,  and E14 as radio SNe.

\vskip 0.2cm
\noindent{\it Transition objects.}\label{se:TC} When the SN blast wave reaches the boundary of the wind-blown bubble it starts to interact with the constant density ISM. This boundary is determined by the balance between wind ram pressure and ISM static pressure. As the  shock wave propagates outward dense shocked ISM gas is accreted,  relativistic particles are efficiently  accelerated and the magnetic field is amplified \citep{BEREZHKO04} causing the source to brighten simultaneously at all radio frequencies. This transition phase lasts approximately until the swept up mass equals the ejecta mass and the source enters its SNR  Sedov phase.   In our data W12 and W25 have rising light-curves at multiple wavelengths and are therefore candidate objects of this type.

A similar increase in flux density at all radio wavelengths has  been observed in SN1987A in the Large Magellanic Cloud \citep[LMC;][]{ZANARDO10}.  
Even though similar to the radio brightening expected in sources in which the blast wave starts to interact with the dense ISM, in the case of SN1987A it is thought that the SN blast wave has started to interact with a dense ring of gas emitted by the progenitor star $20,000$~years before the explosion \citep{BURROWS95}. Hydrodynamic modeling suggests that the dense ring was emitted during the merger of two stars \citep{MORRIS07}.
It is possible that the potential ``transition sources''  in Arp220 also arise from a similar mechanism. Although such merger events must be rare in most galaxies they may be more common in the dense stellar environment of Arp220.

\begin{figure*}[htp]
  \centering
   \includegraphics[width=0.9\hsize]{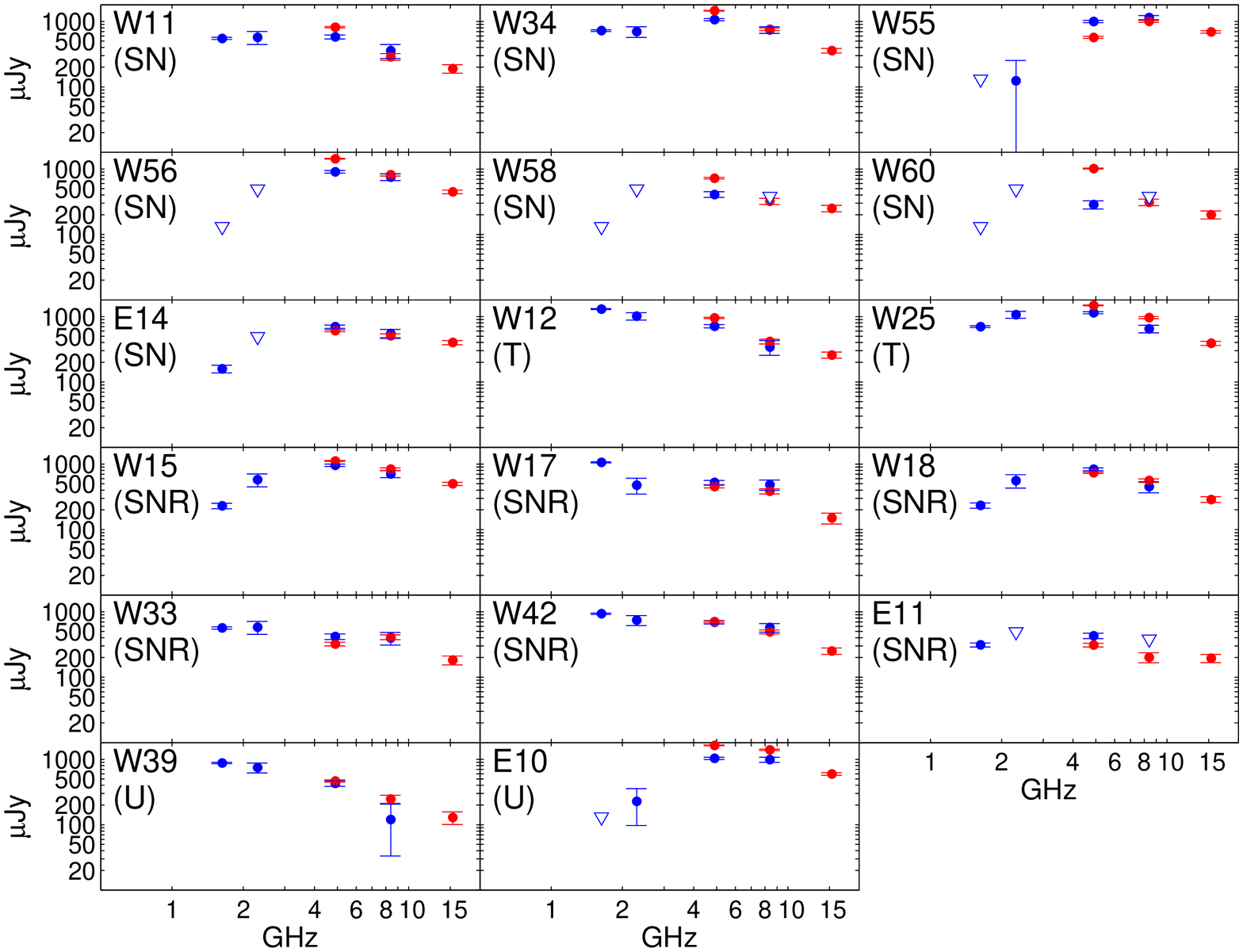}
  \caption{Spectra for the sources detected at 2~cm and 3.6~cm. Blue points are from measurements taken in time period 2006.02--2006.43 and red points in time period 2006.91--2008.44.
   Error bars are plotted at $\pm1\sigma$. In case of non detections upper limits are shown as
  triangles. At 18~cm (1.7~GHz), 13~cm (2.3~GHz), and 3.6~cm (8.4~GHz) upper limits are shown at upper limits of  $6\sigma$, $3.8\sigma$, and $4.4\sigma$
  respectively \citep[see Section~3.1 of][]{PARRA07}.
  The blue data points at 13~cm and at 3.6~cm were taken
  from simultaneous observations in BP129. The blue points at 6~cm (5~GHz) are taken from new reduction of BP129 observations. The new 6~cm, 3.6~cm and 2~cm (15~GHz) points are plotted in red. The 18~cm data are from
  unpublished experiment GD21A (epoch 2006.43). The spectra are grouped according to the classification scheme discussed in
  Section~\ref{se:source_id} where SN = supernova, T = transition object, SNR = supernova remnant, and U = unclassified. For unresolved sources we adopt the 
  measured peak brightness per beam as the best estimate of the total flux density to plot while for resolved sources we plot the integrated flux density within 
  a tight box surrounding the source.
  }
 \label{fig:spectra}
  \end{figure*}

\vskip 0.2cm
\noindent{\it Supernova remnants.}\label{se:SNR}
When the shock created in an SN explosion has swept up an 
ISM mass equal to the  ejecta mass the source enters the Sedov phase of SNR evolution.
In the early part of this phase the total energy in relativistic particles stays approximately 
constant \citep{BEREZHKO04}. In models where the internal shell magnetic field comes from  shock 
compressed external ISM field \citep{VANDERLAAN62,THOMPSON09} the internal magnetic field
is close to constant as the source expands and hence the radio luminosity is also  constant. In contrast for models in which
the magnetic field is internally amplified \citep{BEREZHKO04}, the magnetic energy density
decreases and so does the radio luminosity. A constant or falling  flux density  at all wavelengths is an expected  signature of a radio SNR.  In addition, since the SNR shell is propagating through mainly neutral ISM local free--free absorption effects should be small and  
the radio spectrum close to a power-law. Although free--free absorption  from foreground ionized ISM can occur, this is likely to happen
only at frequencies $\leq2~\mathrm{GHz}$ \citep{PARRA07}. In the Appendix 
we classify sources W15, W17, W18, W33, W42, and E11 as likely SNRs based on the above expected
properties. 
Most of these sources show relatively flat ($\alpha< 0.5$) spectra from 18~cm to 2~cm.
Such spectral indices are unusual but not unprecedented for SNRs. Since all sources discussed in this paper must be detected at 3.6~cm and 2~cm, selection effects will bias spectral indices to small values.

\vskip 0.2cm
\noindent{\it Active galactic nuclei (AGNs).}\label{se:agn}
It has been argued \citep{DOWNECK07} that a 
supermassive black hole and an AGN are present within the western nucleus of Arp220.
Radio emission from an AGN is likely present at some levels and so is potentially detectable.
We would expect 
such an AGN associated compact source  to be long-lived, but perhaps randomly variable 
in intensity. If as expected the radio emission arises within a jet then the spectrum should be 
flat (being generated from the superposition of synchrotron self-absorbed components
peaking at different frequencies along the jet).  Inspection of Figure~\ref{fig:west_map} shows that the
3.6~cm images of the flat spectrum sources W33 and W42 could be consistent with
a jet morphology. Both sources are long-lived and the first shows strong 18~cm variability.
More observations, searching for structure and short-term variability, are needed to 
confirm the presence or absence  of an  AGN in Arp220.

\vskip 0.2cm
\noindent{\it Unclassified sources.}\label{se:unclass}
Two of the sources, described in detail in the Appendix, namely, W39 and E10  have 
proved impossible to classify because they show declining luminosity at long wavelengths 
and increasing luminosity at short wavelengths, a behavior not predicted for any of the classes
described above. More data are required on these sources.

\subsection{Flat Spectrum SNRs}\label{se:source_flat}

It was noted in Section~\ref{se:source_id}  that the sources identified as SNRs mostly have flatter spectra than expected in standard
models (with $\alpha = 0.5 - 0.7$).
Three alternative models could explain the origin of such flat spectra with spectral indices $\alpha \leq 0.5$. In one such model \cite{SCHLICKEISER89} argue that Fermi acceleration in strongly magnetized plasma flattens the electron injection spectrum ($\gamma \leq 2$) which consequently flattens the synchrotron emission spectrum giving $\alpha \leq 0.5$. In another model the intrinsic spectral index is $\alpha \simeq 0.5$ in the SNR shell but the overall spectrum appears flatter due to spatially variable free--free absorption. The integration over the whole SNR of local spectra with turnovers due to free--free absorption happening at different frequencies produces a flatter global spectrum. In the last model the presence of a plerion-like component with a flat or inverted spectrum in the center of the SNR results in an overall radio spectrum with spectral index $\alpha \leq 0.5$. Recent VLBI observations of SN1986J \citep{BIETBART08} show the emergence of a new radio component in the center of the expanding radio shell. The new component shows an inverted radio spectrum contrasting with the power law plus free--free absorption turnover spectrum of the shell; the net result of this is to flatten the integrated spectrum. It is interesting that within Arp220 the spectrum of source W33 might be ``double humped''  (see Figure~\ref{fig:spectra}) showing a peak at $\sim$2~GHz and another one at $\sim$8~GHz. This is consistent with SN1986J like objects. Adding further to this interpretation, it is observed that both W33 and W42 have significantly smaller size at 2~cm than at 3.6~cm  (see Section~\ref{se:size_res}) which would be consistent with a compact central source being more dominant at higher frequency.

\subsection{Comparison of SNe and SNRs Sizes}\label{se:size_lum}

The top panel of Figure~\ref{fig:s_d} plots the  detected sources in the luminosity-diameter plane. This 
figure shows a clear difference in size between 
 sources classified as SNRs (blue symbols) and  SNe (red symbols). All six of the detected SNRs are resolved with  diameter
 $>0.27~\mathrm{pc}$ while all the  SNe (except E14 which has size 0.30~pc)   have sizes $<0.2~\mathrm{pc}$.
 In  normal galactic disks it has been argued  \citep{CHOMWIL09} that the 
 most compact/luminous SNR observed is determined by the age of the youngest SNR expected given 
 the star formation rate. This mechanism cannot however explain the minimum size  of SNRs in Arp220 because 
 given the high predicted  SN rate  ($4\pm2~\mathrm{yr^{-1}}$) the above model predicts a much smaller limit than is 
 observed.
 A physical  minimum size is  in fact expected because SNRs should first ``switch on'' and become 
 luminous just as they enter their Sedov phase, which occurs when the swept up ISM mass equals the ejecta mass.
 The radius at which a source enters the Sedov phase therefore depends on the  ISM  density 
and the ejecta mass.  \citet{SCOVILLE97} estimate in the western nucleus a  mean molecular density of $1.5 \times 10^{4}~\mathrm{cm^{-3}}$; more
recently \citet{SAKAMOTO08} estimate from gas dynamics a mean total mass density of $2\times10^3~M_\odot~\mathrm{pc^{-3}}$ implying a gas molecular number density $<4\times10^{4}~\mathrm{cm^{-3}}$.
Adopting a molecular ISM density of $10^{4}~\mathrm{cm^{-3}}$ a minimum SNR diameter of 0.3~pc is expected, close to that observed. 
This cutoff size for SNRs goes only as the one third power of the assumed ejecta mass 
and external density and so is only weakly dependent on these quantities. Despite this the sharpness of the observed minimum cutoff in  size suggests that the observed Arp220 SNRs are not embedded in densities  $>10^{5}~\mathrm{cm^{-3}}$. This is  either because such high density gas has low volume filling factor in Arp220 or because SNRs become inefficient  radio sources \citep{WHEELER80} when embedded in such high-density regions.

In addition to a {\it minimum} size for  SNRs  we expect a {\it maximum} size for SNe set by the size of their wind-blown bubble.
The predicted sizes of such bubbles are a function of the external ISM pressure, progenitor mass-loss rate and wind velocity \citep[see Equation~2 of][]{PARRA07}.
The ISM pressure in Arp220 is estimated to be  $10^{7}~\mathrm{K~cm^{-3}}$ 
\citep{DOPITA05}. The wind mass-loss rate and speed are expected to vary significantly between different 
progenitors.  \citet{WEILER_REVIEW} estimates however that for the most powerful radio SNe (mostly Type IIn) mass-loss 
rates are  $10^{-4}~M_\odot~\mathrm{yr^{-1}}$ and velocities $10~\mathrm{km~s^{-1}}$. Adopting these parameters gives
an estimated wind-blown bubble diameter of 0.4~pc consistent with the upper limits on SNe sizes and
comparable in size to our SNR diameters. It should be noted however that there is considerable uncertainty 
in the expected wind-blown bubble size since only the ratio of mass-loss rate and wind velocity is constrained
by VLBI SN observations; and for a fixed ratio of these two quantities the bubble size is linearly dependent on 
the assumed wind velocity, a quantity which is poorly known.  The fact that  most of our observed SNe have size limits significantly 
smaller than 0.4~pc suggests that wind speeds could be  $5~\mathrm{km~s^{-1}}$ or less.

In Section~\ref{se:source_id} we discuss the possibility of SN/SNR ``transition 
objects'' where interaction between the blast wave and the ISM has started but the amount of mass accreted is still
less than the ejecta mass. Such objects would be expected to be intermediate in size between SNe and SNRs. 
The two transition object candidates we have,  W12 and W25, have estimated
diameters  $<0.21~\mathrm{pc}$ and $<0.22~\mathrm{pc}$, respectively, which  is consistent with a progenitor with a somewhat  lower mass-loss rate/wind velocity
or a slightly higher ISM  pressure than assumed above. 

It is notable that all of the SN candidates except one lie close together in the luminosity--size plane (see
Figure~\ref{fig:s_d}). All these positions would be reached by the evolutionary tracks of  SNe more radio luminous by approximately a factor of two
than  the well-studied Type IIn SN1986J  (shown by the thick black line marked 86J in Figure~\ref{fig:s_d}). One SN classified source  (E14)  is 
however exceptional since  it lies at a position in the luminosity--size plane more  characteristic of SNRs. Such an intense   
SN  source  which is still highly luminous at late times when it has a large diameter  (and hence likely peaked  at an even higher luminosity),
would not however be  unprecedented.  The radio supernova SN1982aa and  gamma ray bursts/SNe 
SN1998bw and  SN2003w had 6~cm peak luminosities nearly 10 times larger than  SN1986J \citep{CHEVALIER06P}
and their evolutionary tracks would be consistent with the position of E14 in  the luminosity--size plane.

  \subsection{Source Expansion Velocities}\label{se:velocities}

\vskip 0.2cm
\noindent{\it Supernovae.}\label{se:velocities_SNe}
   The majority of the SN identified sources have insufficient data to estimate their explosion 
  dates and hence set limits on their expansion velocities. Three sources
  E14, W11, and W34  have however been detected at several 18~cm epochs 
  showing  close to linearly increasing flux 
  densities with time (Batejat et al. 2012, in preparation).   Comparing to equivalent  portions  of the fitted 18~cm light-curves
  for well-sampled radio SNe of similar luminosities, it seems that  reasonable estimates of explosion 
  dates (within accuracies of one year or so) can be made by linearly extrapolating the 18~cm light-curves down to zero flux 
  density and then subtracting a year.  Applied to E14, W11, and W34, this algorithm gives ages, at the time of the 2~cm
  and 3.6~cm observations,  of 7, 6, and 6 years respectively.  E14 is resolved at both 2~cm and 3.6~cm and  
  W11 is provisionally resolved at 2~cm, allowing to measure their sizes with error bars. The size measurement of W34 we consider unreliable and cannot be used to estimate an expansion velocity (see Section~\ref{se:size_res}). Combining sizes and age estimates for the other two SNe we find that the E14 expansion velocity is in the range $13,700 - 28,000~\mathrm{km~s^{-1}}$ (90\% confidence) and the W11 expansion velocity is in the  range $7300 - 15,200~\mathrm{km~s^{-1}}$. For the latter object the range is consistent with the 
  canonical value of $10,000~\mathrm{km~s^{-1}}$ for a normal  Type II SN. The velocity for E14 is somewhat larger than this 
  but would not be exceptional.  For instance VLBI observations of SN1993J  \citep{MARCAIDE97} 
  show it having a radio shell expansion velocity of $15,000~\mathrm{km~s^{-1}}$ at early times. Also \citet{BRUNTHALER10} have recently measured an expansion velocity of $\sim 23,000~\mathrm{km s^{-1}}$ for SN2008iz in M82.
  
 \vskip 0.2cm
\noindent{\it Supernovae remnants.}\label{se:velocities_SNRs}
 Of the six detected SNRs with measured sizes, four (W17, W18, W33, and W42) 
 have been known since  the original discovery observations of \citet{SMITH98} made in 1994. The remaining two sources, W15 and
 E11, have stable 18~cm light-curves and spectra  consistent with an SNR origin (see the Appendix) but with 
 luminosities below the detection limit of the \citet{SMITH98} observations; these sources were very likely also present 
 in 1994 but were not detectable. It seems highly probable that all the SNR identified sources are at least 12 years old.  In fact 18~cm light-curve monitoring  
 \citep[see Figure~3 in][]{ROVILOS05}  shows their light-curves declining relatively slowly so they are likely considerably 
 older than this. \citet{LONSDALE06} estimate ages of several decades. Our recent reanalysis of all the 18~cm 
 monitoring data (Batejat et al. 2012, in preparation) agrees with this conclusion. Taking a rough estimate of their ages as 20~years and taking 
 the mean diameter for this group of sources (0.31~pc) gives average expansion speeds since explosion of $5000~\mathrm{km~s^{-1}}$. 
 Such speeds are consistent with those expected  for SNRs just entering the Sedov phase when the swept up ISM mass 
 equals that of the ejecta so that strong deceleration is occurring.

\begin{deluxetable*}{cccccc}
\centering
\tablecolumns{6}
\tabletypesize{\scriptsize}
\tablewidth{0pt}

\tablecaption{Magnetic Field and Energy Estimates}

\tablehead{\colhead{Source} & \colhead{$\alpha$} & \colhead{Diameter} &
\colhead{$B$ Rev} &  \colhead{$E$ Part} &  \colhead{$E$ Part, DW} \\ 
\colhead{} & \colhead{} & \colhead{(pc)} & \colhead{(mG)} & \colhead{($\times10^{49}$~erg)} &  \colhead{($\times10^{49}$~erg)} \\
\colhead{(1)}&
\colhead{(2)}&
\colhead{(3)}&
\colhead{(4)}&
\colhead{(5)}&
\colhead{(6)}
}

\startdata

W15 & 0.72 & 0.28 & 23.4 &  3.3 & 3.5\\
W17 & 0.23 & 0.27 & 46.7 &  12.0 & 3.1\\
W18 & 0.62 & 0.38 & 19.9 &  6.0 & 4.8\\
W33 & \nodata & 0.29 & 18.0 & 2.2 & 3.9\\
W42 & 0.24 & 0.31 & 49.7 & 20.4 & 4.4\\
E11 & \nodata & 0.36 & 15.4 & 3.1 & 5.2\\
E14 & 0.72 & 0.30 & 20.9 & 3.3 & 4.0\\

\enddata
\tablecomments{Col. (1): source name. Col. (2): spectral index taken from \citet{PARRA07}. The revised equipartition formula is not valid for spectral indices $\leq0.5$. In cases where a source has $\alpha\leq0.5$, i.e., sources W17 and W42, $\alpha$ is taken as 0.51. In cases where no spectral index information is available, i.e., sources W33 and E11, $\alpha$ is taken as 0.7. Col. (3): source diameter taken from Table \ref{ta:sim_results}. Col. (4): magnetic field derived using the revised equipartition formula of \citet{BECK05} adjusted to a magnetic field to particle energy ratio of 0.01. ${\bf K}$, the ratio of number densities of cosmic-ray protons and electrons per particle energy interval within the energy range traced by the observed synchrotron emission is taken as 100. A shell inner to outer radius of 0.8 and a synchrotron filling factor of 10\% are assumed. Col. (5): associated total energy in relativistic particles. Col. (6): total energy in relativistic particles estimated assuming energy density balance with ram pressure, expansion velocity from \citet{DRAINE91}, a molecular number density of $10^4~\mathrm{cm^{-3}}$, and a synchrotron filling factor of 10\%.}
\label{ta:BK_out}
\end{deluxetable*}

\subsection{Source Magnetic Fields and Energetics} \label{se:b_fields}

Given estimates of source sizes and synchrotron emission flux densities it is possible to work out minimum energies and equipartition magnetic fields. Most models of SNR evolution predict however that SNRs are far from equipartition and that energies are particle dominated. The \citet{BEREZHKO04} model assumes a 1\% ratio of magnetic field to relativistic particle energy density. For the seven sources  which are well resolved at two bands (comprising six SNR candidates and one SN candidate) we give in  Table~\ref{ta:BK_out} estimates of magnetic fields and energies calculated assuming the revised magnetic field - relativistic particle equipartition expression of \citet{BECK05} adjusted to a 1\% magnetic field to particle energy density ratio assumption. We assume a spherical shell geometry with a  ratio of inner to outer radius of 0.8, an outer shell diameter given in Table~\ref{ta:sim_results} and an internal volume filling factor for radio emission of 10\%.
We also assume a ratio of proton to electron number density at fixed energy ${\bf K} = 100$ which according to \citet{BECK05} is valid for young SNRs.
The results give magnetic fields in the range $\sim15$--$50~\mathrm{mG}$ and total energies in relativistic particles between 2\% and 20\% of the expected $10^{51}~\mathrm{erg}$ kinetic energy of a typical SN. This is consistent with estimates given by \citet{LACKI10} for total particle energy fractions of relativistic particles needed to explain the FIR--radio correlation in compact starbursts via calorimeter models.

Consistency with the  \citet{BEREZHKO04} model can be checked by comparing the particle energy in an SNR with the prediction that it should be \mbox{$E_\mathrm{part} = \rho_\mathrm{ISM} v^2 V_\mathrm{shell}$}, where $\rho_\mathrm{ISM}$ is the average density of the ISM, $v$ is the SNR expansion velocity and $V_\mathrm{shell}$ is the synchrotron emitting volume of the SNR. This calculation assumes that ram pressure balances shell internal pressure which in turn is dominated by relativistic particles. Consistent with the latter assumption there is increasing observational evidence for a large/dominant fractional  relativistic particle energy density in young SNRs \citep{PATNAUDE09,  BEREZHKO09}. In Column 6 of Table \ref{ta:BK_out}, we give  the estimated relativistic particle density given by the above ram pressure balance formula. A source-emitting volume $V_\mathrm{shell}$ was calculated from the measured source outer diameter assuming a shell with ratio of inner to outer radius of $0.8$ with synchrotron filling factor of 10\%. The ram pressure $\rho_\mathrm{ISM} v^2$ was calculated from  velocities given by \citet{DRAINE91} for SNRs of the size observed in a medium of density $10^4~\mathrm{cm^{-3}}$; note 
however that this estimate is only weakly dependant on the assumed ISM density since in the Sedov phase the predicted expansion velocity scales as $\rho^{-0.5}$ for fixed diameter. As expected during the Sedov phase  the estimated particle energies in Column 6 of Table~\ref{ta:BK_out} are close to a constant. We find that these estimated relativistic particle energies are in good agreement with those in Column 5 derived  from the measured synchrotron flux densities assuming a 1\% ratio of magnetic to particle energy density.

Estimates of source magnetic fields are given in Column 4 of Table~\ref{ta:BK_out}, as derived from the measured source radio luminosities and sizes assuming a 1\% magnetic 
to relativistic particle energy density ratio and the source geometry and synchrotron volume filling factor described above. \citet{THOMPSON09} estimate somewhat 
smaller field values,  averaging 9~mG for the sources listed in Table~\ref{ta:BK_out}.
These published estimates were made without the benefit of knowing the  source sizes by assuming that 1\% of the total kinetic energy of $10^{51}~\mathrm{erg}$ goes into relativistic electrons and finding the magnetic field that gives the observed source
radio synchrotron luminosity ($\nu L_{\nu}$). It was argued that these relatively modest fields could be produced by shock compression of ISM fields of a few mG; fields sufficiently strong to explain the IR-radio correlation via calorimeter models, even in the presence of inverse Compton energy losses off ambient starlight photons \citep{THOMPSON09}.
This compression model predicts however that SNRs have constant radio intensities until the electron synchrotron energy loss time ($\geq$ 200 years at 5~GHz for the sources in Arp220). Most  observable sources would then be older than 100 years old and therefore be  larger than 0.7~pc in diameter in a $10^4~\mathrm{cm^{-3}}$ density medium \citep{DRAINE91}, a factor two greater than their measured sizes. An additional consequence of such a model, given the high SN rate in Arp220,  is that many more SNR sources than the six candidates presented in this paper would be expected to be detected (order of a few hundred). The low number of observed SNRs in Arp220 combined with their small sizes argue against models in which SNR magnetic fields are generated from compressed ISM magnetic fields.

 \subsection{The  SNR Luminosity-Size Relation}\label{se:flux_d}

A correlation between the radio 
surface brightness~($\Sigma$) of SNRs and their diameters~($D$) has long been 
claimed \citep{SHKLOVSKII60}. Such a correlation  can alternatively be cast as 
one between radio luminosity~($L$)
and diameter~($D$),  a formulation  which  removes the  implicit $D^{-2}$ correlation 
automatically induced by plotting $\Sigma$.  Both the reality and slope of the correlation has
been a controversial topic given the likelihood of strong selection effects,  both  in our Galaxy \citep{GREEN05} and 
(to a lesser extent) in external  galaxies \citep{URO05a}. It has also been claimed that if the  $\Sigma$--$D$ 
and $L$--$D$  correlations exist, they are secondary correlations due to correlations with the ISM density \citep{BERK86,BANDIERA10}.  
Finally, it has been claimed  \citep{ARBUTINA05}  that different correlations exist for  the luminous SNRs in galactic  
dense clouds and most  observed extragalactic SNRs compared to galactic SNRs in low  density regions.

\begin{figure*}[htp]
  \centering 
  \includegraphics[width=0.9\hsize]{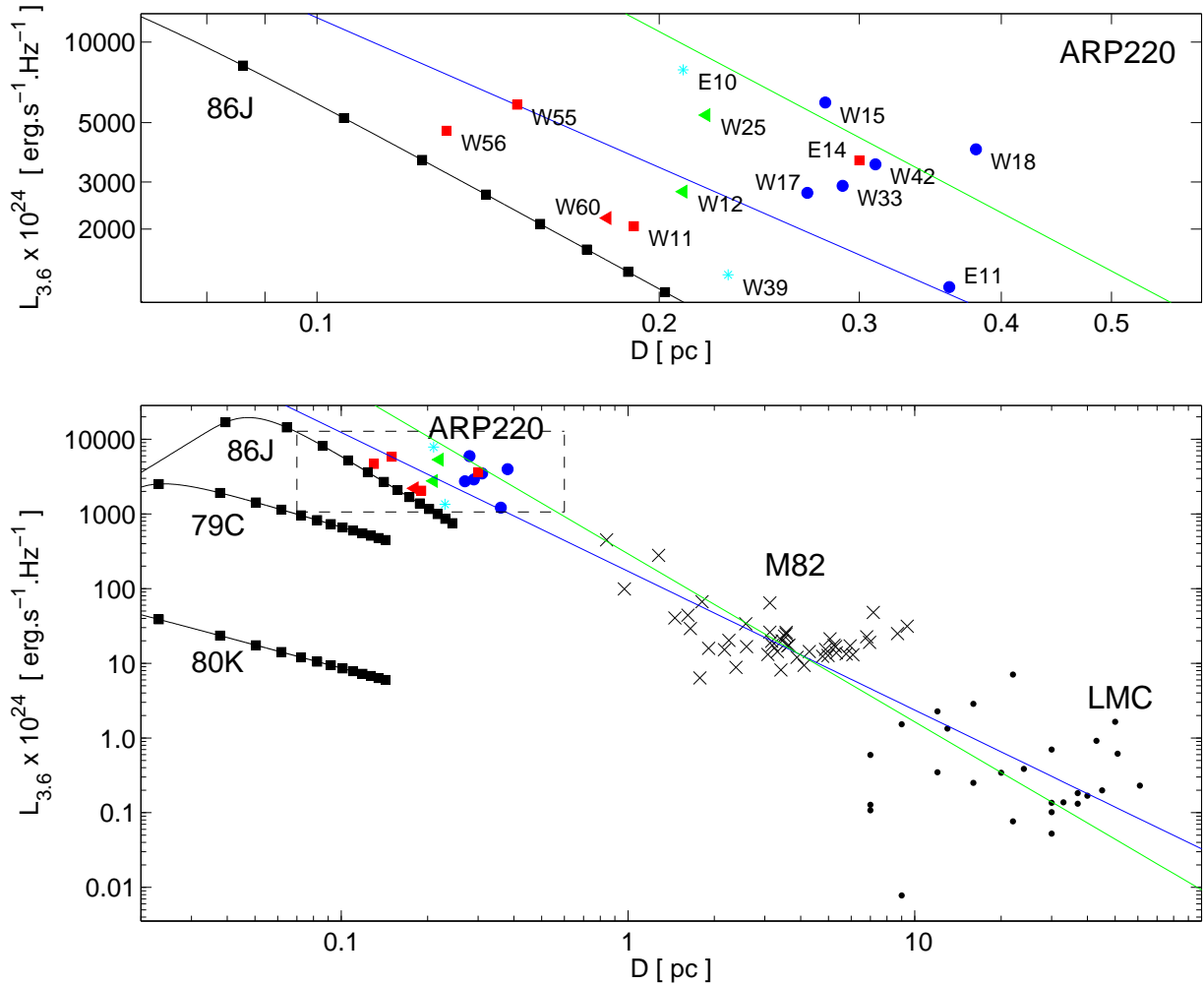}
  \caption{\footnotesize Illustration of the
    3.6~cm flux densities and sizes of observed SNe and SNRs.
    The bottom panel plots sources in Arp220 and SNRs in M82 and the LMC.
    The top panel is a close up view of the dashed rectangular region shown in
    the bottom panel showing only Arp220 sources.
    Red is used for sources identified as SNe, blue
    for sources identified as SNRs, green for possible transition objects, and cyan
    for unclassified sources.
    The SN tracks for SN1986J
    (Type~IIn), SN1979C (Type~IIL), and SN1980K (Type~IIL) were
    produced using the light-curve fits given in \citet{WEILER_REVIEW}
    combined with the deceleration parameter from \citet{BIET02} for
    SN1986J and assuming free expansion at \E{4}~\kms\ for both
    SN1980K and SN1979C. The square markers along each track indicate
    time evolution and are 1~year apart. 
    Black crosses and dots show SNRs in M82 and the LMC.
    W34 and W58 are omitted as their size estimates are not
    considered dependable. In cases where a point is denoted by a triangle pointing left
    this signifies that the size estimate is an upper limit. In both panels the green line is the 
    theoretical line of \citet{BEREZHKO04} of slope $-9/4$ while the blue line is the best fit through the
    SNR luminosities and sizes for all three galaxies.
    The data for the SNRs in M82 are taken from \citet{HUANG94} and the data for
    the SNRs in the LMC from \citet[][see Section~\ref{se:flux_d}]{MILLS84}.}
 \label{fig:s_d}
\end{figure*}

 We can test the claimed $L$--$D$ correlation for powerful SNRs 
 by comparing the  diameters and luminosities  of the SNRs in Arp220 with those  found 
in  M82 and the LMC (see  bottom panel of Figure~\ref{fig:s_d}).
The data for 45 SNRs in M82 used in   Figure~\ref{fig:s_d} are taken from \citet{HUANG94} who use the FWHM of
a Gaussian fit as an estimate of source diameter. To make these estimates compatible with our Arp220 
estimates we have converted this to a dispersion and then 
to the diameter of a shell model with ratio of inner to outer radius 0.8. The data 
for the LMC are taken from \citet{MILLS84}. These authors obtained sizes by taking two models,
one comprising a thin circular ring and another a
uniform circular disk and fitting to the half power response along the major and minor axes. An
average of the two values is chosen and is expected to give a reasonable approximation to the 
shell size as defined for Arp220. 

Figure~\ref{fig:s_d} clearly shows that the data for Arp220 are 
to first order consistent  with an extension of the $L$--$D$ correlation to very small size SNRs. More 
quantitatively  an unweighted  least-squares fit to all SNR points gives the best-fitting  relation $L \propto D^{-1.9}$ (drawn in blue
in Figure~\ref{fig:s_d}). This gradient is however mostly determined by the M82 and LMC data because of the larger 
number of  SNRs plotted for these galaxies.
  
Models for radio SNR evolution such as those presented by \citet{HUANG94} and
\citet{BEREZHKO04} posit constant ratios between relativistic  particles and  field energy densities and roughly constant energy 
in  relativistic particles during the Sedov phase. Individual SNRs evolve along the luminosity-diameter correlation 
during their Sedov phase on tracks which are independent of the 
external density  $n_\mathrm{ISM}$ \citep{BEREZHKO04}.\footnote{Despite this direct independence of $L$ on $n_\mathrm{ISM}$ these 
models can explain that the most  luminous SNRs are found in host nuclei with 
high densities such as Arp220 because (1) such nuclei have the highest star formation rates and hence the
youngest, smallest diameter, and so most luminous  SNRs \citep{CHOMWIL09} and because (2) there is a 
minimum ``switch on''
size for Sedov phase SNRs, which decreases as density increases, increasing the maximum SNR luminosity.} 
Assuming an optically thin radio synchrotron spectral index of $0.5$, these models predict \mbox{$L \propto D^{-9/4} E^{7/2}$}, where $E$\ is the SN kinetic energy. Assuming that all SNe have close to the same energy then we expect  \mbox{$L \propto D^{-9/4}=D^{-2.25}$}.
Assuming this dependence and fitting the constant of proportionality we obtain the green line in Figure~\ref{fig:s_d}. This theoretical correlation fits the observed Arp220 data very well and also, given their large internal scatter, is consistent with the M82 and LMC data.

Some competing models argue that radio SNR luminosity is instead mostly determined by ISM density with source  size being of secondary importance.
In one such model \citep{VANDERLAAN62, THOMPSON09} the internal magnetic fields that control SNR luminosity are compressed ISM fields which themselves increase with ambient density
($B_\mathrm{ISM}\propto n_\mathrm{ISM}^{\alpha}$)  giving rise to a $L$--$n_\mathrm{ISM}$  relation.  A secondary $L$--$D$ relation can however still occur because the diameter of 
typically observed SNRs  also depends on  ISM density.
Specifically in the model of    \citet{VANDERLAAN62} the radio luminosity of an SNR at a given frequency stays roughly constant till a time equal to the electron synchrotron loss time, $t_\mathrm{syn}$ after which it rapidly  declines.  Given  that  SNRs are decelerating  during their Sedov  and subsequent pressure-driven snow-plow phases  it follows that most observed radio SNRs in a flux-limited sample will have sizes close to their sizes at the synchrotron loss time $t_\mathrm{syn}$.  For most sources plotted in Figure~\ref{fig:s_d} the predicted $t_\mathrm{syn}$ is comparable or larger than the age and which they enter the snow-plow phase for which \mbox{$D \propto n_\mathrm{ISM}^{-1/7} t_\mathrm{syn}^{2/7}$} \citep{MCKEE77};  substituting $t_\mathrm{syn}\propto B_\mathrm{ISM}^{-3/2}$ and $B_\mathrm{ISM}\propto n_\mathrm{ISM}^{\alpha}$ a  $D$--$n_\mathrm{ISM}$ relationship  results. Combining this with the predicted $L$--$n_\mathrm{ISM}$  relationship  noted earlier we find that we get  an $L$--$D$ relationship with power-law exponent  of $-2.25$ if the power-law exponent linking ISM magnetic field and density has a value $\alpha = 0.6$.

\citet{THOMPSON09}  assuming the  \citet{VANDERLAAN62} SNR  model have estimated $B_\mathrm{ISM}$ in a number of galaxies and have compared this to estimates of  ISM surface density $\Sigma$ finding the empirical relation  \citep[Equation~(6)]{THOMPSON09}  \mbox{$B \propto \Sigma^{0.55}$}.
If one assumes a constant gas scale height then \mbox{$n \propto \Sigma$} and \mbox{$B_\mathrm{ISM} \propto n^{0.55}$}; similar to the value of the exponent required above to explain the slope of the $L$--$D$ relation.  Despite this
consistency there is no physical explanation of why the exponent value is $0.55$.\footnote{\citet{THOMPSON09} instead argue that physically \mbox{$B_\mathrm{ISM} \propto n$} might be expected based on equipartition between the magnetic field energy density and the energy density required to balance disk gravity}
In contrast the model of \citet{BEREZHKO04} directly predicts the slope of the observed $L$--$D$ from the model's assumed physics. Additionally in Section~\ref{se:b_fields} we gave arguments based on source energetics, sizes and numbers  that the \citet{BEREZHKO04} model applies in Arp220. The fact that the \citet{BEREZHKO04} model naturally explains the $L-D$ relation {\it between} galaxies 
and that all galaxies fall on the same $L$--$D$ relation argues that this is the only mechanism needed and that it also  applies in M82 and the LMC. The \citet{BEREZHKO04} model in normal galaxies is also favored by the work on SNR luminosity functions by \citet{CHOMWIL09}.

A strong test which can differentiate between density-dependent models and the 
 \citet{BEREZHKO04} model would be to look for the time variations in luminosity predicted to occur only in 
 the latter model as  individual sources expand and move along the $L$--$D$ correlation.  Given the small 
sizes of the SNR sources in Arp220 quite large flux variations are expected over
relatively short times. For instance if our SNRs  are expanding at $3\,000~\mathrm{km~s^{-1}}$ just as they 
enter the Sedov phase, then for  a diameter $\sim0.3~\mathrm{pc}$ they 
increase  in diameter  by $2\%~\mathrm{yr^{-1}}$  which from \citet{BEREZHKO04} implies  a  flux density decrease of 
 $4.8\%~\mathrm{yr^{-1}}$.  \citet{ROVILOS05}, who analyzed five epochs of 18~cm data  spanning 
 5.6 years, were able to rule out factor of  two variations over that period, as expected
from SNe models, but not variations of the amplitude predicted above. We are presently analyzing 
 nine 18~cm data sets over a longer time period  to see  if we can detect 
 the predicted flux density decline.

  \subsection{ISM Magnetic Fields in Arp220}\label{se:ISM_fields}
 
In Sections~\ref{se:b_fields} and \ref{se:flux_d} we argue that in SNRs internally generated magnetic fields dominate over compressed ISM  magnetic fields both in Arp220 and other galaxies. The compressed ISM magnetic field contribution must therefore be significantly less than our total estimated internally generated field, i.e., $f\,B_\mathrm{ISM} \ll B_\mathrm{SNR}$, with $f$ ranging from 3 to 6 for young SNRs \citep{VOLK02}. Taking the median $B_\mathrm{SNR}$ from Table~\ref{ta:BK_out} we get an upper limit on $B_\mathrm{ISM}$ ranging from $3.5~\mathrm{mG}$ to $7~\mathrm{mG}$ in Arp220. This is consistent with the estimates of magnetic fields from OH maser Zeeman splitting \citep[$0.7$--$4.7~\mathrm{mG}$,][]{ROBISHAW08}.
For comparison \citet{THOMPSON09} estimate a minimum ISM magnetic field of $\sim 1~\mathrm{mG}$ in order that synchrotron emissivity dominates over inverse Compton losses. \citet{THOMPSON09} also estimate an upper limit of 20~mG based on the argument that the magnetic field is dynamically lower than gravity.


\section{Conclusions} \label{se:conclusions}

The main conclusions of this paper are as follows.
\vskip 0.3cm

\noindent\textbf{1.} We have detected two new sources at both  2~cm and 3.6~cm wavelength in the western nucleus of Arp220, namely W58 and W60.

\noindent\textbf{2.} We have resolved for the first time, 11 of the 17 detected sources
at 2~cm, 8 at 3.6~cm, and 7 at both 2~cm and 3.6~cm. 

\noindent\textbf{3.} We confirm the claim of \citet{PARRA07} that the compact radio sources in Arp220 mostly comprise a
mixed population of SNe and SNRs. Two sources are candidate SN/SNR transition objects. A few sources remain difficult to classify and may be 
AGN components. Ongoing VLBI monitoring at 6~cm should shed further light on
the nature of these sources.

\noindent\textbf{4.} The sources resolved at both wavelengths (all but one of which are SNRs) have diameters in the range 0.27--0.38~pc with mean 0.31~pc. 
In comparison the SNe are, except in one case (E14), all unresolved. The observed size boundary between SNe and SNRs is consistent with an ISM density of $\sim10^4~\mathrm{cm^{-3}}$.

\noindent\textbf{5.} Combining source sizes with source ages enables us to calculate upper limits for source expansion velocities averaged over their lifetime. We find expansion
velocities \mbox{$<$~30,000~$\mathrm{km~s^{-1}}$} for the SNe W11 and E14 and $\sim5000~\mathrm{km~s^{-1}}$ for the SNRs.

\noindent\textbf{6.} We argue that magnetic fields in the SNRs of Arp220 are internally generated \citep{BEREZHKO04} and not dominated by compressed ISM magnetic fields \citep{VANDERLAAN62}. This interpretation is supported by the fact that the  particle energy density estimated from synchrotron fluxes and sizes, assuming a 1\% ratio of magnetic field to relativistic particle energy density ratio, is equal to the ram pressure energy density, just as predicted by \citet{BEREZHKO04}. Furthermore the relativistic particle total energies are consistent with values required to fit the FIR-radio correlation \citep{LACKI10}. In contrast the compressed field model is  inconsistent with the low number of observed SNRs and their small measured diameters.

\noindent\textbf{7.} The observed  SNR radio luminosity as a function of diameter for sources in Arp220, M82 and the LMC is in good agreement with the predicted  
relation  $L_{\nu} \propto D^{-2.25}$ as derived assuming internally generated magnetic fields in SNR shells \citep{BEREZHKO04}.

\noindent\textbf{8.} Based on our conclusion that in SNRs internal magnetic fields dominate over compressed ISM magnetic fields we estimate an upper limit 
of 7~mG for the ISM magnetic field in Arp220. This is consistent with other estimates and limits.

\acknowledgements{This research was partly supported by the EU Framework 6 Marie Curie Early Stage Training programme under contract number MEST-CT-2005-19669 ``ESTRELA''. J.C. acknowledges funding from a Swedish VR research grant. The European VLBI Network is a joint facility of European, Chinese, South African and
other radio astronomy institutes funded by their national research councils. The National Radio Astronomy Observatory is a facility of the National Science Foundation
operated under cooperative agreement by Associated Universities, Inc. The authors thank the staff at NRAO and
JIVE for their diligent correlation of the data and the referee whose comments helped improve the paper.}


\section*{Appendix}

This appendix assigns source type classifications (mostly SN or SNR) to the high frequency detected sources in Arp220.
In Section~\ref{se:source_id} we discuss the 
expected temporal and spectral properties of each of  the possible source classes.
The most critical information for classification is the variability properties of each source.
Concentrating on recent high sensitivity data we find that source variability properties fall naturally into four different groups which we term ``rapidly rising'', ``possibly rising'', ``stable'', and ``possibly declining''. 
The term ``rapidly rising'' is used in the individual source descriptions below at 18~cm and 6~cm for a source whose flux density has risen more than $3\sigma$ and 30\%, between
two high sensitivity epochs over $\sim 2.5$~years (comparing the two 6~cm epochs in Table~\ref{ta:source_fluxes} and the  2003.85 and 2006.44 epochs at 18~cm). 
The same term  is used at 3.6~cm for a source whose flux density has risen by more than $3\sigma$ and 15\% over the 0.89~years between the two epochs.
The term ``possibly rising'' is used for a source whose flux density has risen by more than $1\sigma$ and the term ``possibly declining'' 
is used for a source whose flux density has declined by more than $1\sigma$. The term ``stable'' is used for any variation 
in flux density lower than $1\sigma$.
\\

\noindent W11 - First detected at 18~cm in epoch 2003.85 this source is rapidly rising 
at 18~cm. It is also  rapidly rising at 6~cm and stable at 3.6~cm. Its spectrum peaks
around 3~GHz.  We classify W11 as an SN.\\

\noindent W12 - First detected at 18~cm in epoch 2002.88  this source is rapidly rising at 18~cm and 6~cm. A similar rise rate 
at 3.6~cm  is consistent with the data.  The radio spectrum shows a 
straight power-law spectrum from  18~cm to 3.6~cm.  We classify this source  as a candidate SN/SNR transition object.   \\

\noindent W15 - First detected at 18~cm in epoch 2002.88. \citet{PARRA07} classified this source as ambiguous because
 it was detected in the 18~cm observations GD17A and GD17B  but at levels below the \citet{ROVILOS05} sensitivity limit, meaning that it could be
either a new source or a long-lived stable source not detected earlier because of sensitivity
limitations. Subsequent data, showing a stable 18~cm light-curve, are more consistent with the latter interpretation. However, a possible rise at both 6~cm (16\%) and  3.6~cm (18\%) may  
be in contradiction with an SNR origin. Despite this, considering potential $\sim10\%$ calibration/reconstruction uncertainties, the large size of this object 
(0.28~pc resolved at both 6~cm and 3.6~cm bands) and its spectrum peaking around 3~GHz, we provisionally classify W15 as an SNR.\\

\noindent W17 - A long-lived source discovered by \citet{SMITH98}. Its long-term 18~cm light-curve is stable and its 6~cm and 3.6~cm light-curves are possibly 
declining. Its spectral shape is quite flat but could be interpreted as being double humped with one peak at $\sim1.5~\mathrm{GHz}$ and another at $\sim5~\mathrm{GHz}$. 
We classify W17 as an SNR.\\

\noindent W18 - A long-lived source first discovered by \citet{SMITH98}.  Long term 18~cm flux density 
monitoring  shows a steady decrease of 8\%~$\mathrm{yr^{-1}}$ over 11.6 years.  In more recent shorter wavelengths 
at  6~cm and 3.6~cm the source is classified, respectively, as possibly  declining and  possibly rising.
W18's spectrum turns over around 3~GHz. We provisionally classify W18 as an SNR. This 
classification is  based primarily on its  long-lived nature and declining light-curves at 18~cm and 6~cm.\\

\noindent W25 - First observed at 18~cm in epoch 2002.88. It is rapidly rising at 18~cm, 6~cm and 3.6~cm. Its spectrum peaks around 3~GHz. 
Primarily based on its rapidly  rising light-curves at all frequencies we consider this source to be a candidate SN/SNR transition object.\\

\noindent W33 - A long-lived source first discovered in \citet{SMITH98}. Its 18~cm and 6~cm light-curves are possibly 
declining over $\sim 2.5~\mathrm{years}$ while its 3.6~cm flux density has remained stable over 0.89~years. On average 
its 18~cm flux density has decreased $\sim10\%$~$\mathrm{yr^{-1}}$ since monitoring began in 1994.
The spectrum of W33 is quite flat with hints of two broad peaks at $\sim$2~GHz and $\sim8~\mathrm{GHz}$.
 \citet{PARRA07} suggested that W33 is a possible AGN candidate. Inspection of the 3.6~cm image of W33 does not exclude
the possibility of a jet morphology.  Based on the behavior of its multifrequency light-curves we
provisionally classify W33 as an SNR but an AGN origin is still possible.\\

\noindent W34 - First discovered at 18~cm in epoch 2003.85. This was followed by a rapid rise at 18~cm epochs. 
W34's 6~cm light-curve is also rapidly rising while the source is stable at 3.6~cm. W34's spectrum turns over around  5~GHz.
The light-curves and spectrum are consistent with it being an SN.\\

\noindent W39 - A long lived source detected by \citet{SMITH98}.  Long term monitoring shows a steady decline at 18~cm ($\sim7\%~\mathrm{yr^{-1}}$).
From recent short wavelength data  it is classified as stable at 6~cm and possibly rising at 3.6~cm. The source's spectrum turns over
at $\sim2~\mathrm{GHz}$. Given its unusual multi-frequency variability properties (declining at long wavelengths but stable or rising at 
short wavelengths) we refrain from speculation about the nature of W39 until more data are available.\\

\noindent W42 - A long-lived source first detected by \citet{SMITH98}. It has decreased by less than 8\% in 18~cm flux density 
since 1994 while its recent 6~cm and 3.6~cm light-curves are stable. W42 is the most stable of the long lived sources. The
spectrum is quite flat from 18~cm to 3.6~cm. There is no sign of a low frequency turnover but there is a high-frequency
turn down from 3.6~cm to 2~cm. We conclude that W42 is likely to be a  SNR.\\

\noindent W55 - First detected at high frequency by \citet{PARRA07} in epoch 2006.02.  No 18~cm data are available.
The source's 3.6~cm light-curve is possibly declining, the  6~cm light-curve is rapidly  declining 
and its spectrum peaks at around 8~GHz. 
 We provisionally classify W55 as an SN. \\

\noindent W56 - This source was first discovered by \citet{PARRA07} in epoch 2006.02. No 18~cm data are available. The
source is rapidly rising at 6~cm and stable at 3.6~cm. W56's spectrum turns over around 5~GHz. Considering these properties we
conclude that W56 is an SN.\\

\noindent W58 - A newly detected source in the most recent  2~cm and 3.6~cm observations. At these two wavelengths it  has flux densities of 226 and $322~\uJy$,
respectively. At this 3.6~cm flux density the source would have been less than a 4$\sigma_\mathrm{rms}$ detection in BP129
(epoch 2006.02).   New 6~cm data from GC031A  yield a flux density of $723~\uJy$. Re-reduction of BP129  6~cm data made it possible to clearly detect this source
at epoch 2006.02 with a flux density of $408~\uJy$ \citep[not detected by][]{PARRA07} allowing us to classify it as rapidly rising ($77\%$ increase in $\sim 2.5$~years). No 18~cm or 
3.6~cm light-curve information is available. We classify W58 as an SN.\\

\noindent W60 - A newly detected source in the most recent  2~cm and 3.6~cm observations. 
The source was also detected in the latest 6~cm observation in experiment GC031A (epoch
2008.44) at a flux density of 1016~$\mu$Jy. Re-reduction of BP129 6~cm data made it possible to clearly detect this source at epoch 2006.02 with a
 flux density of 285~$\mu$Jy \citep[not detected by][]{PARRA07} allowing us to classify this source  
  as rapidly rising at  6~cm. We classify W60 as an SN.\\

\noindent E10 - First detected at 18~cm in epoch 2002.88. It showed a decline in the
next epoch and was undetectable in the last two 18~cm epochs. 
The source spectrum peaks at $\sim5~\mathrm{GHz}$. 
This observed fast  low frequency decline is too rapid to be consistent with an SNR and suggests 
instead a rapidly evolving Type Ib/c SN observed after its peak at 18~cm. However 
in contradiction to this interpretation the source is rapidly rising 
at 6~cm and 3.6~cm. We leave E10 unclassified until further data are available.\\

\noindent E11 - First detected in the high sensitivity 18~cm  epoch 2002.88   at a  flux
density below the  detection limit of earlier epochs.  Its 18~cm flux density has
decreased by less than 4\% in subsequent  18~cm  monitoring spanning $\sim 3.5$~years.
It was not detected by \citet{PARRA07} at 6~cm and 3.6~cm. Re-reduction of BP129 6~cm data (epoch 2006.02)
 made it possible to clearly  detect this source with a flux density of 429~$\mu$Jy compared 
to 310 ~$\mu$Jy in experiment GC031A (epoch 2008.44)  allowing us to classify it  
as possibly declining at this wavelength.  Based on its stable 18~cm light-curve and 
possibly declining 6~cm light-curve we classify E11 as an SNR.\\

\noindent E14 - First detected at 18~cm in epoch 2002.88.  In subsequent monitoring 
it shows a rapidly rising 18~cm light-curve  over $\sim 2.5$~years. The source is classified as possibly declining
at 6~cm and stable at 3.6~cm. 
The spectrum shows a turnover frequency  at 2~GHz. We classify E14 as an SN.\\


\bibliographystyle{apj}
\bibliography{ms}

\end{document}